\documentclass{jfm}

\usepackage{graphicx}
\usepackage{natbib}
\usepackage{parskip}
\usepackage{subfigure}

\ifCUPmtlplainloaded \else
  \checkfont{eurm10}
  \iffontfound
    \IfFileExists{upmath.sty}
      {\typeout{^^JFound AMS Euler Roman fonts on the system,
                   using the 'upmath' package.^^J}%
       \usepackage{upmath}}
      {\typeout{^^JFound AMS Euler Roman fonts on the system, but you
                   dont seem to have the}%
       \typeout{'upmath' package installed. JFM.cls can take advantage
                 of these fonts,^^Jif you use 'upmath' package.^^J}%
      }
  \else
  \fi
\fi

\ifCUPmtlplainloaded \else
  \checkfont{msam10}
  \iffontfound
    \IfFileExists{amssymb.sty}
      {\typeout{^^JFound AMS Symbol fonts on the system, using the
                'amssymb' package.^^J}%
       \usepackage{amssymb}%
       \let\le=\leqslant  
         
      }{}
  \fi
\fi

\ifCUPmtlplainloaded \else
  \IfFileExists{amsbsy.sty}
    {\typeout{^^JFound the 'amsbsy' package on the system, using it.^^J}%
     \usepackage{amsbsy}}
    {}
\fi

\newsavebox{\astrutbox}
\sbox{\astrutbox}{\rule[-5pt]{0pt}{20pt}}

\title[Turbulent Particle Pair Diffusion]{On Turbulent Particle Pair Diffusion}

\author[Nadeem A. Malik]%
{N\ls A\ls D\ls E\ls E\ls M\ns A.\ns M\ls A\ls L\ls I\ls K
\thanks{Email: namalik@kfupm.edu.sa,  and nadeem\_malik@cantab.net}}

\affiliation{Department of Mathematics and Statistics,
King Fahd University of Petroleum and Minerals,
P.O. Box 5046, Dhahran 31261, Kingdom of Saudi Arabia\\[\affilskip]}

\pubyear{2010}
\volume{650}
\pagerange{119--126}
\date{?; revised ?; accepted ?. - To be entered by editorial office}
\begin{document}

\maketitle

\begin{abstract}
Richardson's theory of turbulent particle pair diffusion [Richardson, L. F. Proc. Roy. Soc. Lond. A 100, 709–737, 1926], based upon observational data, is equivalent to a locality hypothesis in which the turbulent pair diffusivity $(K)$ scales with the pair separation $(\sigma_l)$ with  a 4/3-power law, $K\sim \sigma_l^{4/3}$. Here, a reappraisal of the 1926 dataset reveals that one of the data-points is from a molecular diffusion context; the remaining data from geophysical turbulence display an unequivocal non-local scaling, $K \sim \sigma_l^{1.564}$. Consequently, the foundations of pair diffusion theory have been re-examined, leading to a new theory based upon the principle that both local and non-local diffusional processes govern pair diffusion in homogeneous turbulence. Through a novel mathematical approach the theory is developed in the context of generalised power law energy spectra, $E(k)\sim k^{-p}$ for $1<p\le 3$, over extended inertial subranges. The theory predicts the scaling, $K(p)\sim \sigma_l^{\gamma_p}$, with $\gamma_p$ intermediate between the purely local and the purely non-local scalings, i.e. $(1+p)/2<\gamma_p\le 2$. A Lagrangian diffusion model, Kinematic Simulations [Kraichnan, R. H., Phys. Fluids 13, 22-31, 1970; Fung et al., J. Fluid Mech. 236, 281-318, 1992], is used to examine the predictions of the new theory all of which are confirmed. The simulations produce the scalings, $K\sim \sigma_l^{1.545}$ to $\sim \sigma_l^{1.570}$, in the accepted range of intermittent turbulence spectra, $E(k)\sim k^{-1.72}$ to $\sim k^{-1.74}$, in close agreement with the revised 1926 dataset. 


\end{abstract}


\linespread{1.5}

\section{Introduction}

Turbulent transport and mixing play an essential role in many natural and industrial processes \cite{Shraiman2000}, in cloud formation \cite{Vaillancourt2000}, in chemical reactors and combustion systems \cite{Pope1994}, and atmospheric and oceanographic turbulence determines the spread of pollutants and biological agents in geophysical flows \cite{Huber2001}, \cite{Berloff2002}, \cite{Wolf2004}, \cite{Joergensen2005}. Concentration fluctuations are often of great importance in such systems and this is related to the separation of nearby fluid particles; turbulent pair diffusion therefore plays a critical role in such systems.
The idea of locality has been fundamental to the theory of turbulent particle pair diffusion since Richardson's pioneering paper, \cite{Richardson1926}, which established turbulent pair diffusion as an important scientific discipline and laid the foundations for a theory of how ensembles of pairs of fluid particles (tracers) initially close together move apart due to the effects of atmospheric winds and turbulence. Richardson argued that as particle pairs separate the rate at which they move apart is affected mostly by eddies of the same scale as the separation distance itself – this is the basis of the locality hypothesis.
Richardson was also motivated by a desire to bring molecular and turbulent pair diffusional processes into a unified picture through the use of a single non-Fickian diffusion equation with scale dependent diffusivity, $K(r)$, where $r$ is the pair separation variable. Assuming homogeneous isotropic turbulence, Richardson posed the problem in 3D in terms of the probability density function (pdf) of the pair separation, $q=q(r,t)$, and assuming the normalization,
$\int_0^{\infty} 4\pi r^2 q(r,t) dr=1$, he suggested the following diffusion equation to describe $q$,
\begin{eqnarray}
  {\partial q\over \partial t} &=& 
  {1\over r^2} {\partial \over \partial r} \left({r^2 K(r){\partial q\over \partial r}}\right)
\end{eqnarray}

The scaling of the pair diffusivity $K$ with the pair separation and what it means for the pair diffusion process is of paramount importance. From observational data of turbulent pair diffusivities collected from different sources, Richardson assumed an approximate fit to the data namely, $K\sim l^{4/3}$. This is equivalent to  $\langle l^2\rangle \sim t^3$, \cite{Obukhov1941}, \cite{Batchelor1952}, often referred to as the Richardson-Obukov $t^3$-regime. $l(t)$ is the pair separation, $t$ is the time, and the angled brackets is the ensemble average over particle pairs.

It is no longer believed that it is possible to unify molecular and turbulent diffusional processes because their physics are fundamentally different; Brownian motion characterizes molecular diffusion, while convective gusts of winds that increase the pair separation in surges characterizes turbulent diffusion \cite{Wilkins1958}, \cite{Fung1992}, \cite{Virant1997}. Nevertheless, the idea of a scale dependent turbulent diffusivity has survived. 

Richardson's assumed 4/3-scaling law is equivalent to a locality hypothesis, according to which for asymptotically large Reynolds number only the energy in eddies whose size is of a similar scale to the pair separation inside the inertial subrange is effective in further increasing the pair separation. Furthermore, Richardson's scaling for the pair diffusion is consistent with \cite{Kolmogorov1941}, (K41); this can be seen from the form of the turbulence energy spectrum in the inertial subrange which is, $E(1/l)\sim \varepsilon^{2/3} (1/l)^{-5/3}$, from which it follows that the pair diffusivity depends only upon $l$ and $\varepsilon$ (the rate of kinetic energy dissipation per unit mass). This leads directly to the locality scaling, $K\sim \varepsilon^{1/3} l^{4/3}$. It is usual to evaluate $K$ at typical values of,  $l$, namely at $\sigma _l=\sqrt{\langle l^2\rangle}$,  so this scaling is replaced by, $K\sim\varepsilon^{1/3} \sigma_l^{4/3}$. The requirement of large Reynolds number, $Re\to \infty$, implies that this scaling is true only inside an asymptotically infinite inertial subrange, $\eta \ll \sigma_l\ll L$, where $L/\eta \to \infty$, and  $\eta$ is the Kolmogorov micro-scale and $L$ is a scale characteristic of energy containing eddies, typically the integral length scale, or the Taylor length scale. In this limit, the pair separation is initially zero, $ l\to 0$, as $t\to 0$.

For finite inertial subrange, there are infra-red and ultra-violet boundary corrections so the inertial subrange still has to be very large in order to avoid these effects and to observe inertial subrange scaling in the pair diffusion. With the 4/3-scaling for $K$, an explicit solution for equation (1.1) for diffusion from a point source with boundary conditions $q(0,t)=q(\infty,t)=0$, can be derived,
\begin{eqnarray}
  q(r,t) &=& 
  {429\over 70} \sqrt{{143\over 2}} \left({1\over \pi \langle r^2 (t)\rangle}\right)^{3/2}  
  \exp⁡{\left({ -\left({1287\over 8\langle r^2 (t)\rangle}\right)^{1/3} }\right)}
\end{eqnarray}

	Turbulence is both scale dependent and time correlated (non-Markovian), and it is not clear whether the pdf in equation (1.2), which describes a local and Markovian process, can accurately represent the turbulent pair diffusion process. Attempts have been made to derive alternative non-Markovian models for pair diffusion, \cite{Falkovich2001}, \cite{Eyink2013}, and this remains a subject of ongoing scientific research. However, this does not affect Richardson's hypothesis of scale dependent diffusivity, which is the main focus of this study.

	It is possible to generalize the scaling for the pair diffusivity to be time dependent and still be consistent with K41 and with locality, \cite{Klafter1987}, \cite{Salazar2009}, such that, 

\begin{eqnarray}
	K &\sim& \varepsilon^a t^b l^c
\end{eqnarray}
for some $a,b$,and $c$. Dimensional consistency then gives, $2a+c=2$ and $3a-b=1$, which leads to $a=1/3-b/3$, and $c=4/3-2b/3$. Thus we obtain, 

\begin{eqnarray}
	K &\sim& \varepsilon^{(1/3  - b/3)} t^b l^{(4/3  - 2b/3)}
\end{eqnarray}
If the further constraint $2b+3c=4$ is satisfied, then this yields, $\langle l^2 \rangle\sim t^3$. Thus, a $t^3$-regime  is not a unique signature for Richardson's 4/3-scaling for the pair diffusivity. However, a time dependent pair diffusivity is hard to justify physically if we assume steady state equilibrium turbulence, because a time dependent diffusivity implies, for $b>0$ that the pair diffusivity at the same separation is ever increasing in time without limit, or for $b<0$ that the pair diffusivity approaches zero with time and the separation process effectively stops. Both cases seem unlikely, and in the ensuing we will restrict the discussion to steady state equilibrium turbulence and consider only the case, $b=0$. However, it is worth remarking that time-dependent diffusivities like equation (1.3) may exist in the context of non-equilibrium turbulence.

	The main aims of this research are two-fold. Firstly, to re-examine the body of evidence available on turbulent pair diffusion especially, if any, from large scale turbulence containing large inertial subranges, Section 2. Particular attention is focussed upon a reappraisal of Richardson's original 1926 dataset, Section 3. Secondly, based upon these findings, a new theory that does not {\em a priori} make the assumption of locality is constructed from which new scaling laws for the pair diffusivity is derived through a novel mathematical method, and compared to the reappraised 1926 dataset, in Section 4. Finally, a Lagrangian diffusion model is used to examine the new theory and the results from the simulations are found to agree closely with the 1926 dataset and with the predictions of the new theory, Section 5. We discuss the significance of these findings and draw conclusions in Section 6.

\section{What is the evidence for locality?}\label{sec2}

Although the general consensus among scientists in the field at the current time is that the collection of observational data, experimental data, and Direct Numerical Simulation, suggests a convergence towards a Richard-Obukov locality scaling, the relatively low Reynolds numbers in the experiments and DNS, and the high error levels, and other assumptions made in collecting the data means that this is by no means a forgone conclusion. As noted by \cite{Salazar2009}, " .. there has not been an experiment that has unequivocally confirmed R-O scaling over a broad-enough range of time and with sufficient accuracy."  
It is not known precisely what size of the inertial subrange is required to observe unequivocally the pair diffusion scaling, but it is widely assumed to be about at least four orders of magnitude or more. Only geophysical turbulent flows, such as in the atmosphere and in the oceans, can produce such extended inertial subranges. 

A number of claims of observing locality scaling, including the existence of an approximate $t^3$-regime in geophysical flows, have been made by \cite{Tatarski1960}, \cite{Wilkins1958}, \cite{Sullivan1971}, and \cite{Morel1974}. More recent observations include \cite{Julian1977} in the atmosphere, and \cite{Lacasce2003}, and \cite{Ollitrault2005} in the oceans. But high error levels and other assumptions (such as two-dimensionality) made in these observations means that decisive conclusions cannot be drawn from them regarding pair diffusion theories.

‘Direct Numerical Simulations (DNS) is inconclusive at the current time because it does not produce a big enough inertial subrange in order to be able to test pair diffusion laws reliably. For example, \cite{Ishihara2009} perform a DNS with $4096^3$, at Taylor scale based Reynolds numbers $R_\lambda\approx 1200$ showing an approximate inertial subrange energy spectrum over a very short range of just 40.  Other DNS of particle pair studies at low Reynolds numbers are, \cite{Yeung1994} at $R_\lambda=90$, \cite{Boffetta2002} at $R_\lambda=200$, \cite{Ishihara2002} at $R_\lambda=283$, \cite{Yeung2004} at $R_\lambda=230$, \cite{Sawford2008} at $R_\lambda=650$. \cite{Scatamacchia2012} at $R_\lambda=300$. See also \cite{Bitane2012}, and \cite{Biferale2014}. The maximum separation of time scales between the integral time scale and the Kolmogorov times scale observed to date in DNS is about a factor of, 100. This is still about two orders of magnitude smaller than the minimum size required to adequately test inertial subrange pair diffusion scalings. For 2D turbulence, see \cite{Jullien1999}, \cite{Boffetta2000}.

Particle Tracking Velocimetry (PTV) laboratory experiments \cite{Maas1993}, \cite{Malik1993} have been providing pair diffusion statistics at low to moderate Reynolds numbers. Like DNS, these Reynolds numbers are too small to reliably test pair diffusion laws. \cite{Virant1997} obtain pair diffusion measurements from PTV.  More recently, \cite{Berg2006} obtained measurements in a water tank at $R_\lambda=172$, and \cite{Bourgoin2006} and \cite{Ouellete2006} tracked hundreds of particles at high temporal resolution at $R_\lambda=815$. Although higher resolution tracking experiments using high-energy physics methods have been performed for single particle trajectories \cite{Laporta2001}, they have not yet been applied to particle pair studies.

When intermittency is accounted for, the scaling in fully developed turbulence should in fact be slightly greater than the R-O scaling. This will be discussed further in Discussions and Conclusions, Section 6.

\section{A reappraisal of the 1926 observational dataset}

\cite{Richardson1926} reported in the Proceedings of the Royal Society of London data on turbulent diffusivities collected from different sources, which is reproduced here with a brief description in Table 1.
He plotted the turbulent diffusivity against the pair separation in log-log scale, shown as the red and black filled circles in Figure 1. Motivated by an attempt to unify pair diffusional processes across all possible scales, he assumed the scaling  $K\sim l^{4/3}$ as a reasonable fit (Fig. 1, dotted blue line). This is not the least squares line of best fit to the data, it is just a theoretical approximation to the data. The actual line of best fit is, $K\sim l^{1.248}$ (Fig. 1, solid red line).

\cite{Richardson1948} and \cite{Stommel1949}, commenting on diffusion of floats in the sea noted that their new measurements were roughly consistent with the 1926 data, i.e. $K\sim l^{4/3}$. 
At the end of their paper they wrote, "Note added in proof. – After this manuscript was submitted the writers have read two unpublished manuscripts by C. L. von Weisaecker and W. Heisenberg in which the problem of turbulence for large Reynolds number is treated deductively with the result that they arrive at the 4/3 law. The agreement between von Weisaecker and Heisenberg’s deduction and our quite independent induction is a confirmation of both", see \cite{Benzi2011}. 
But they also observed that, "any power between $\sim l^{1.3}$ and $\sim l^{1.5}$ would be a tolerable fit to the data".  Thus, as early as 1948, it was noted that $K\sim l^{4/3}$ was only a rough fit, which makes it all the more surprising that an alternative theory for pair diffusion has not been developed till now.

\begin{table}
\begin{center}
\def~{\hphantom{0}}
  \begin{tabular}{l l cc}
 Datum    &            &Diffusivity $K$  & Scale $l$\\      
 Number  &{\hfill Source\hfill}   & [$cm^2/s$]   & [$cm$]\\       
   &&& \\     
 $N1$ &Molecular diffusion of oxygen in to ntirogen [1] &$1.7\times 10^{-1}$ &$5\times 10^{-2}$ \\[-3pt]
 $N2$ &Anemometers 9 m above the ground [2] &$3.2\times 10^{3}$ &$1.5\times 10^{3}$ \\[-3pt]
 $N3$ &Anemometers 21-305 m above the ground [3] &$1.2\times 10^{5}$ &$1.4\times 10^{4}$ \\[-3pt]
 $N4$ &Pilot balloons 100-800 m above the ground [4,5] &$6\times 10^{4}$ &$5\times 10^{4}$ \\[-3pt]
 $N5$ &Tracks of balloons in the atmosphere [6,7] &$1\times 10^{8}$ &$2\times 10^{6}$ \\[-3pt]
 $N6$ &Volcano ash [6,7]  &$5\times 10^{8}$ &$5\times 10^{6}$ \\[-3pt]
 $N7$ &Diffusion from cyclones [8] &$1\times 10^{11}$ &$1\times 10^{8}$ \\[-3pt]
   &&& \\     
 \end{tabular}

{\linespread{1.2}
 \caption{Datum number, source, the turbulent diffusivity, $(K)$, and the length scale, $(l)$. 
References:
[1] \cite{Kay1982}, [2] \cite{Schmidt1917}, [3] \cite{Akerblom1908}, 
[4] \cite{Taylor1915}, [5] \cite{Hesselberg1915}, [6] \cite{Richardson1922}, 
[7] \cite{Richardson1925}, [8] \cite{Defant1921}\protect}\label{tab01}
}
 \end{center}
 \end{table}

\begin{figure}
\begin{center}
\mbox{\subfigure{\includegraphics[width=12cm]{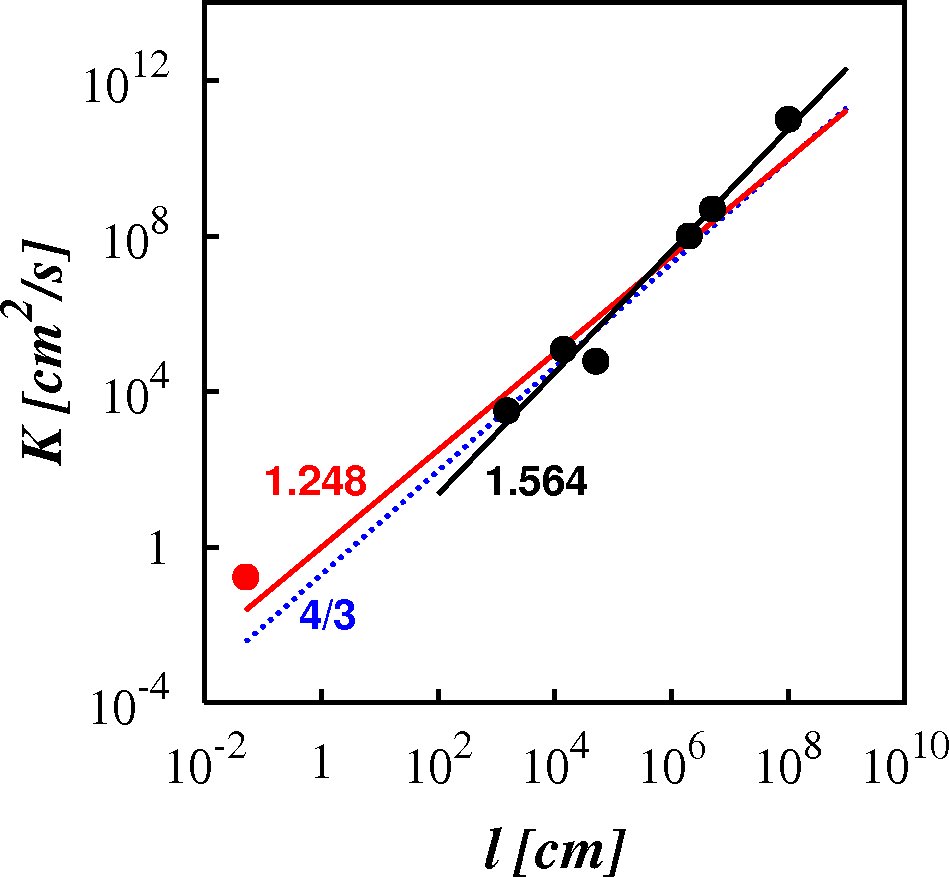}}}

{\linespread{1.2}
\caption{
The turbulent pair diffusivity against separation, as $\log(K)$ against $\log(l)$. The symbols are the observational data reported by Richardson in 1926 (Table 1). The red filled circle (N1) is from of the molecular diffusion of oxygen into nitrogen. The black filled circles (N2-N7) are from geophysical settings. The dotted blue line is Richardson's assumed locality scaling, $K\sim \sigma_l^{4/3}$. The solid red line is the least square line of best fit to the entire 1926 dataset (N1-N7), $K\sim \sigma_l^{1.248}$. The solid black line is the least square line of best fit to the revised dataset (N2-N7), $K\sim \sigma_l^{1.564}$, and the coefficient of determination is, $R^2=0.97$.\protect}\label{fig01}
}
\end{center}
\end{figure}

However, there is a more fundamental problem with the data. Data-point number N1 in Table 1 (Fig. 1, red filled circle) is not from turbulence measurements at all, rather it is from studies of the molecular diffusion of oxygen into nitrogen and whose length scale is stated to be of the order of $10^{-2}$ cm. At such a small scale it cannot be turbulent, and it cannot contain an extended inertial subrange (even if it were turbulent). This data-point must therefore be disregarded as an outlier in the current investigation which strictly demands the existence of turbulence with an extended inertial subrange over several orders of magnitude in scale.

The remaining six data-points (Fig. 1, black filled circles) are sound, coming from geophysical turbulence settings and certainly containing extended inertial subranges. The line of best fit to this new improved dataset (N2 to N7) displays an unequivocal non-local scaling, $K\sim l^{1.564}$ (Fig. 1, solid black line). The coefficient of determination, $R^2=0.97$, is very close to 1, meaning that all the data-points lie very close to line of best fit, as is obvious from Figure 1. This implies that the true scaling must be very close this scaling power, probably between $\sim l^{1.5}$ and  $\sim l^{1.6}$, which is outside of the bounds noted by Richardson \& Stommel, and far from the locality scaling.

The importance of this new finding cannot be under-estimated because it is clear that the 1926 dataset itself implies that turbulent pair diffusional process cannot be purely local in nature, and therefore non-local diffusional processes cannot be ignored {\em a priori} in any general theory of pair diffusion, which is where we turn to in the next section.


\section{A new theory based upon locality and non-locality}

To develop a theory based upon the fundamentally new physical picture that non-local as well as local diffusional processes may play a significant role in the turbulent pair diffusion process, a new mathematical approach is now constructed from which new scaling laws for the turbulent pair diffusivity can be derived.  Like Richardson, the focus here is on the diffusivity, which is the most important quantity in the problem; the mean square separation $\langle l^2\rangle$ is related to the pair diffusivity by the exact relation, $K=0.5d\langle l^2\rangle/dt$, and with a general scaling for the diffusivity, $K\sim\sigma_l^\gamma$, there is a the corresponding scaling for $\langle l^2\rangle\sim t^\chi$. For steady-state equilibrium turbulence $\chi$ is given by the exact relation,

\begin{eqnarray}
	\chi &= &{1\over {1-\gamma/2}},  \qquad  1<p< 3,
\end{eqnarray}

As such  $\langle l^2\rangle$ does not provide any additional information.  We therefore focus our analysis mainly upon $K$, and refer to $\langle l^2\rangle$ only where needed in the ensuing analysis and discussions.

To elucidate the role of local and non-local diffusional processes in turbulent pair diffusion, here turbulence with generalized energy spectra, $E(k)\sim k^{-p}$, $k_1\le k\le k_\eta$ with extended inertial subrange, $k_\eta/k_1\gg 1$, in the power range $1<p\le 3$, is considered. The term ‘wavenumber’ will be used to denote both the vector, k, as well as its magnitude, $k=|k|$, and the context will make it clear. Such spectra are routinely used in pair diffusion studies; it helps in understanding the changes in the balance of diffusional processes as the turbulence spectra changes in a way that looking at just Kolmogorov turbulence alone cannot do. In the current work, generalised spectra will play an important mathematical role in validating the new theory for pair diffusion.

The pair diffusivity is now a function of  $p$,  i.e. $K(p)$. For some other quantities the subscript “$p$” will be used to denote dependence upon the energy spectrum.

\subsection{The statement of the problem}

The problem is to determine the pair diffusivity, $K=\langle {\bf l\cdot v}\rangle$, of an ensemble of pairs of fluid particles in a field of homogeneous turbulence containing an extended inertial subrange. The particles in a pair are located at ${\bf x}_1 (t)$ and ${\bf x}_2 (t)$ at time $t$, the pair displacement vector is ${\bf l}(t)={\bf x}_2 (t)-{\bf x}_1 (t)$, and the pair separation is ${\bf l}(t)=\sqrt{l_1^2+l_2^2+l_3^2 }=|{\bf x}_2 (t)- {\bf x}_1 (t)|$. The initial separation at some earlier time, $t_0$, is denoted by $l_0=|{\bf x}_2 (t_0 )- {\bf x}_1 (t_0)|$. The turbulent velocity field is, ${\bf u}({\bf x},t)$, and the particle velocities at time t are, respectively, ${\bf u}_1 (t)={\bf u}({\bf x}_1 (t),t)$ and ${\bf u}_2 (t)={\bf u}({\bf x}_2 (t),t)$, and the pair relative velocity is ${\bf v}(l)={\bf u}_2 (t)-{\bf u}_1 (t)$.

As earlier mentioned, Richardson's theory holds strictly in the limit of infinite inertial subrange, implying that $l_0\to 0$, as $t\to 0$ . For a large but finite inertial subrange and $l_0\not=0$, short time corrections need to be made to the theory, an issue addressed by \cite{Batchelor1952}. It is well known that there exists an initial ballistic regime for very short times due to the high correlation with the initial conditions, which leads to,  $\langle l^2\rangle \sim (t-t_0)^2$ – the so-called Batchelor regime. At much larger times, when the pair separation is of the order of the integral length scale, their motions become independent and the pair diffusion collapses to twice the one-particle Taylor diffusion, $\langle l^2\rangle \to 2 \langle x^2\rangle \sim t$. These two regimes are well understood and will not be considered any further as the interest here is in the scalings inside the inertial subrange.

In turbulent pair diffusion studies, it is often assumed that the initial pair separation is smaller than the Kolmogorov micro-scale, $l_0<\eta$, although this is not a formal requirement of the theory. The particles will diffuse apart and eventually decorrelate with the initial conditions, they will 'forget' their initial conditions, $(l_0,t_0 )$, as Batchelor put it, after some travel time, $t_{l_0}$, when the pair is inside the inertial subrange of turbulent motions. The transition from the Batchelor regime to the explosive inertial subrange regime occurs on a time scale of the order to the eddy turnover time at scale, $\tau_0 (l_0)$. If $l_0$ is already inside the inertial subrange then, $t_{l_0}\approx \tau_0 \sim \varepsilon^{1/3} l_0^{1/3}$.

Without loss of generality, it will be assumed that, $t_0=0.$ If the pair loses dependency on the initial conditions after some travel time $t_{l_0}$ when the separation, $\delta=\sqrt{\langle l^2\rangle (t_{l_0})}\ll L$, is well inside the inertial subrange, then it is from $t=t_{l_0}$ that inertial range scaling is assumed to apply.  $t_{l_0}$ need not be precisely determined, so long as it is close to when initial conditions have been 'forgotten'. It has been found that taking $t_{l_0}$ to be the time when the pair ensemble distance is equal to the Kolmogorov scale, $\delta=\eta$, gives good numerical results \cite{Fung1992}. 

When considering generalized power law spectra in simulations, there will be occasions when a start with initial separation greater than the Kolmogorov scale is needed, $\eta<l_0\ll L$; in these cases, at large times after release but when the pair is still inside the inertial subrange, it is reasonable to approximate with $\delta=l_0$ and $\tau_0\approx 0$,  because $t\gg \tau_0$; the numerical results in Section 6 will justify this approximation. On the basis of the locality hypothesis, the turbulent pair diffusivity is given from dimensional arguments, \cite{Richardson1926}, \cite{Obukhov1941}, \cite{Batchelor1952}, as

\begin{eqnarray}
  K &\sim& \varepsilon^{1/3} \sigma_l^{4/3}, \qquad Max(\eta,\delta)\ll \sigma_l\ll L, \quad t\gg t_{l_0}
\end{eqnarray}

For statistically steady equilibrium turbulence this is equivalent to, $ \langle l^2\rangle\sim t^3$ \cite{Obukhov1941}.

\subsection{The mathematical framework}

The pair relative velocity is, ${\bf v}=d{\bf l}/dt$, and the pair diffusivity is defined as the ensemble average of the scalar product of ${\bf v}$ with ${\bf l}$,
\begin{eqnarray}
	K &=& \langle {\bf l\cdot v} \rangle	
\end{eqnarray}

The interest in this research is in the scaling laws for the pair diffusivity, so a policy of suppressing constants wherever possible will be followed in the rest of the paper.

For homogeneous, isotropic, incompressible, reflectional and statistically stationary turbulence, the Fourier expression for the velocity field ${\bf u}$ is, \cite{Batchelor1953},

\begin{eqnarray}
	{\bf u}({\bf x}) &=& \int {\bf A(k)}  \exp⁡{(i{\bf k\cdot x})} d^3 {\bf k}	
\end{eqnarray}

where ${\bf A(k)}$ is the Fourier transform of the flow field, ${\bf k}$ is the associated wavenumber. The quantity of interest is the relative velocity ${\bf v}$ across a finite displacement ${\bf l}$, 

\begin{eqnarray}
	{\bf v}({\bf l}) &=& {\bf u}({\bf x}_2 )-{\bf u}({\bf x}_1).
\end{eqnarray}

Using equation (4.4) this is gives,

\begin{eqnarray}
 {\bf v}({\bf l}) &=& \int {\bf A(k)} [\exp⁡{(i{\bf k\cdot l})}-1 ]  \exp⁡{(i{\bf k\cdot x_1})} d^3{\bf k}.
\end{eqnarray}

Taking the scalar product of ${\bf v}$ with ${\bf l}$ and then the ensemble average $\langle\cdot\rangle$ over particle pairs yields an expression for $\langle {\bf l\cdot v} \rangle$;  in diffusion studies, it is usual to assume that the Lagrangian ensemble scales with this quantity, see \cite{Thomson2005} for example. Thus we obtain a scaling for the pair diffusivity,

\begin{eqnarray}
  K &\sim& \langle{\bf l\cdot v}\rangle  
  \sim \int \langle{\bf (l\cdot A)}[\exp⁡{(i{\bf k\cdot l})}-1 ]\exp⁡{(i{\bf k\cdot x_1})}\rangle d^3{\bf k}.
\end{eqnarray}

Because of homogeneity, the ensemble average removes the factor $\exp⁡(i{\bf k\cdot x_1})$ without altering the scaling upon $l$. This gives,

\begin{eqnarray}
   K(l) &\sim& \int \langle{\bf (l\cdot A)}[\exp⁡{(i{\bf k\cdot l})}-1 ]\rangle d^3{\bf k}.
\end{eqnarray}

Let $k$ denote the magnitude of the wavenumber, $k=|{\bf k}|$. Let $k_l=1/l$ be the wavenumber associated with the pair separation. As before, the scaling, $l\sim \sigma_l$, will be assumed throughout this work, so that $k_l\sim 1/\sigma_l$, where $\sigma_l^2=\langle l^2 \rangle$. 

It is tempting to approximate the integrand in (4.8) by expanding the exponential term such that, $\exp⁡{(i{\bf k\cdot l})}-1\approx i{\bf k\cdot l}$. However, such an expansion is accurate only for low wavenumbers  $k\ll k_l$ where, $|{\bf k\cdot l}|\ll 1$. For local wavenumbers $k\approx k_l$ where $|{\bf k\cdot l}|\approx 1$, this expansion is approximate and implies some corrections. For high wavenumbers $k\gg k_l$ where $|{\bf k\cdot l}|\gg 1$, this expansion is not accurate.

\subsection{Three diffusional processes and uncertainty}

A fundamental physical assumption about the nature of the diffusional processes that are occurring in the system is made here, namely it is assumed that there exist three independent diffusional processes that potentially contribute to the pair diffusion process as a whole, each process acting from its own range of wavenumbers relative to the inverse pair separation wavenumber $k_l$. Partitioning of the spectrum in to different ranges on such a physical basis is often used in turbulence theory, see for example the work on wall turbulence by \cite{Perry2005}.

For a given pair separation, $\sigma_l$, the three physical processes operate, respectively, from the scales of motion that are smaller than $\sigma_l$, from the scales that are local to $\sigma_l$, and from the scales that are non-local to $\sigma_l$. The associated frequencies are, $\omega(k)\propto \sqrt{k^3 E(k)}$, according to the usual assumption that the frequencies scale with the inverse turnover time of the eddy at wavenumber $k$. The local eddy frequency is, $\omega_l\propto \sqrt{k_l^3 E(k_l)}$, and the local eddy turnover time is, $T_l\sim 1/\omega_l$. 

On this physical basis, the integral in equation (4.8) is partitioned in to a sum of three integrals over different wavenumber ranges which are defined by,

\noindent $s$: the small scales such that $k\gg k_l$, and $|{\bf k\cdot l}|\gg 1$, and the associated frequencies are much larger than $\omega_l$,  $\omega(k)\gg \omega_l$.

\noindent $l$: the local scales such that $k\approx k_l$, and $|{\bf k\cdot l}|\approx 1$, and the associated frequencies are of the same order as $\omega_l$,  $\omega(k)\approx \omega_l$.

\noindent $nl$: the non-local scales such that $k\ll k_l$, and $|{\bf k\cdot l}|\ll 1$, and the associated frequencies are much smaller than $\omega_l$,  $\omega(k)\ll \omega_l$.

With this partitioning, (4.8) becomes

\begin{eqnarray}
  K &\sim& \left({\int_{nl}+ \int_l+ \int_s}\right) \langle ({\bf l\cdot A})
       (\exp⁡{(i{\bf k∙l})-1})\rangle d^3 {\bf k}.
\end{eqnarray}

which is rephrased as,

\begin{eqnarray}
	K &\sim& K^{nl}+K^l+K^s
\end{eqnarray}

There is uncertainty in defining the 'size' of a turbulent eddy, a problem which is inherent to the nature of turbulence, \cite{Tennekes1972}. Consequently, it is not possible to define the precise cut-offs between the local and non-local scales and between the local and small scales of motion. 

Although this appears to be an intractable problem, such uncertainty is a defining characteristic of turbulence and is implicit even in the locality hypothesis where a lack of precise definition of the size of a local eddy does not prevent scaling laws from being obtained. Likewise, as the ensuing analysis will show, the general scaling for the relative diffusivity and some asymptotic results can also be obtained in the more general case considered here.

\subsection{The physics of the small scales of motion}

A simplification can be made with respect to the small scales of motion whose contribution to the diffusion process is,

\begin{eqnarray}
	K^s &\sim& \int_s \langle{\bf (l\cdot A)} (\exp⁡{(i{\bf k\cdot l})}-1)\rangle d^3{\bf k}
\end{eqnarray}

There is no need to evaluate this integral directly, because the net ensemble effect can be assessed on physical grounds alone. $K^s$ is the integral over high wavenumbers and represents the contribution from scales of turbulent motion which are much smaller than the pair separation, i.e., from $k\gg k_l$. The energy contained in these scales is very small if the energy spectrum decreases as k increases, such as an inverse power law of the type $E(k)\sim k^{-p}$, with $p>1$. 

Furthermore, these small scales are associated with unsteadiness of high frequencies, $\omega^s\gg \omega_l$. Statistically, these high frequency motions induce random and rapid changes in the direction and magnitude of the pair displacement vector. 

Overall, the changes in the statistics of the pair separation induced by the action of the high frequency, small scale, low energy, random turbulent velocity fluctuations is expected to be extremely small. It is reasonable to assume that the net ensemble effect of the high wavenumbers in the pair diffusion process is correspondingly small, i.e., $K^s\ll Max(K^l,K^{nl})$. $K^s$ will therefore be neglected henceforth.

\subsection{The physics of the local and non-local scales of motion}

With the effect of the small scale contributions eliminated, the simplified expression for the pair diffusivity is,

\begin{eqnarray}
  K &\sim& K^{nl}+K^l \\ &\sim&
 \int_{nl} \langle{\bf (l\cdot A)} (\exp⁡{(i{\bf k\cdot l})}-1)\rangle d^3{\bf k} +
 \int_l      \langle{\bf (l\cdot A)} (\exp⁡{(i{\bf k\cdot l})}-1)\rangle d^3{\bf k}
\end{eqnarray}

The expansion of the exponential in the integrand to leading order is accurate only in the non-local range where $|{\bf k\cdot l}|\ll 1$. In the local range where $|{\bf k\cdot l}|\approx 1$, such an expansion is only approximately true. Such local eddies are moderately unsteady with frequencies that are of the same order of magnitude as, $\omega_l$. The effect on the local diffusion process is assumed to be likewise moderate, without killing it entirely.
  
Hence, it is assumed that the ensemble effect of the unsteadiness in the local wavenumbers is to quantitatively reduce the magnitude of the local diffusivity, $K^l$, but without altering its overall scaling behaviour. Then, the expansion, $\exp(i{\bf k\cdot l})-1\approx i{\bf k\cdot l}$, can be used in (4.13) but with the magnitude of $K^l$ reduced by some factor,  $F_l\lesssim 1$, which is smaller than unity, but not too small. Then (4.13) becomes,

\begin{eqnarray}
   K &\sim& \int_{nl}\langle {\bf (l\cdot A)}{(i\bf k\cdot l)}\rangle  d^3 {\bf k} +
      F_l \int_l \langle{\bf (l\cdot A)}{(i\bf k\cdot l)}\rangle d^3 {\bf k}
\end{eqnarray}

$F_l=F_l (p,R_l,C)$ is not expected to be a universal constant because it will depend upon various parameters, like $p$, and $R_l=k_l/k_1$ which is the size of the inertial subrange relative to the particle pair separation, and also implicitly upon the size of an eddy in wavenumber space, $C$ (to be defined later). After absorbing constants, the integrands in (4.14) become,

\begin{eqnarray}
	\langle l^2 |{\bf A}||{\bf k}|  \cos⁡(\alpha)  \cos⁡(\beta)  \rangle
\end{eqnarray}
		
where $\alpha$ is the angle between $\bf l$ and $\bf A$, and $\beta$ is the angle between $\bf l$ and $\bf k$. For isotropic random fields averaging (4.15) over all directions, again, does not affect the scaling behaviour. $\alpha$ and $\beta$ are not uniformly distributed in all direction, it is well known that ${\bf l}$ aligns preferentially in the positive strain directions, and this ensures that the ensemble average above is non-zero. 

Retaining the angled brackets $\langle \cdot \rangle$ to include averaging over all directions, (4.14) with (4.15) then simplify to,

\begin{eqnarray}
	K(l) &\sim& \int\int_{nl} \langle l^2 ak\rangle  dk dA(k) +F_l\int\int_l \langle l^2 ak\rangle  dk dA(k)	
\end{eqnarray}

where $a=|{\bf A}|$, and $dA(k)$ is the element of surface area at radius $k$ in wavenumber space. If the closure, $\langle l^2 ak\rangle \sim \langle l^2\rangle \langle ak\rangle$, is assumed, then upon integrating over the surface area this integral becomes

\begin{eqnarray}
	K(l) &\sim&   \left({ \int_{nl }\langle ak\rangle dk + 
                     F_l \int_{nl }\langle ak\rangle dk }\right) \langle l^2\rangle.	
\end{eqnarray}

Because $k$ and $a$ are magnitudes, then $\langle ak\rangle\not=0$ even though the vectors ${\bf k}$ and ${\bf A}$ are orthogonal.

$\int \langle a^2\rangle dA(k)$ is the energy density per unit wavenumber averaged over all directions, \cite{Batchelor1953}, and scales like $\sim E(k)/k$. If the closure, $\int \langle ak\rangle dA(k)\sim k\sqrt{(∫\langle a^2 \rangle dA(k))}$, is assumed then (4.17) becomes,

\begin{eqnarray}
	K(l) &\sim& \left( {\int_{nl} \sqrt{kE(k)} dk + F_l \int_{nl} \sqrt{kE(k)} dk }\right)
             \langle l^2 \rangle	
\end{eqnarray}

As a further check, this can also be derived as follows. The velocity variance from the scales $k$ to $k+dk$ is $E(k)dk$, and the variance of velocity gradient is $k^2 E(k)dk$. The particle pair velocity variance is, $\sim \langle l^2 \rangle k^2 E(k)dk$. The time scale of eddies of wavenumber $k$ is $1/\sqrt{k^3 E(k)}$. So the incremental contribution to the diffusivity from these scales is the pair velocity variance times the time scale, $dK\sim \langle l^2 \rangle \sqrt{kE(k)} dk$, which leads to (4.18). 

To make further progress the actual form of the turbulence spectrum must be specified. In this work, the focus is upon high Reynolds number turbulence which contains an extended inertial subrange. For pair diffusion statistics, the form of the energy spectrum in the large energy containing scales is not important. Such scales do, however, sweep the inertial range eddies which must be modeled. This is implemented by working in the swept frame of reference by setting the spectrum in the large energy scales to zero, $E(k)=0$ for $k<k_1$, and assuming an inverse power-law energy spectrum in the inertial subrange,

\begin{eqnarray}
	E(k) &=& c\varepsilon^{2/3} L^{5/3-p} k^{-p}, \qquad k_1<k<k_\eta, \quad 1<p\le 3
\end{eqnarray}
 
where $c$ is a constant. A large length scale $L$ is necessary for dimensional consistency. $L$ scales with some length scale that is characteristic of the large energy scales, such as the integral length scale, or the Taylor length scale.

With the spectrum in (4.19), and with $\sigma_l^2 =\langle l^2 \rangle$, equation (4.18) becomes,

\begin{eqnarray}
	K(p) &\sim& \varepsilon^{1/3} L^{(5/3-p)/2} 
      \left({ \int_{nl} k^{(1-p)/2} dk + F_l \int_l k^{(1-p)/2} dk}\right) \sigma_l^2	
\end{eqnarray}

This is the most general expression for $K$ that can be derived from the present analysis without any {\em a priori} assumption regarding locality, other than the partitioning of the spectrum in to local and non-local ranges.

To check the effectiveness of the mathematical approach adopted here in deriving equation (4.20), the locality limit from this expression must first be validated against Richardson's locality hypothesis for which there is a known theory.

\subsection{Validation: the locality limit}

The assumption of locality means that the non-local (first) term on the right hand side in equation (4.20) is ignored. To evaluate the remaining local integral, some cut-off wavenumber $k_*$, such that $k_1<k_*<k_l$, which defines the size of a local eddy in wavenumber space is assumed. Locality is thus assumed valid in the wavenumber range $k_*< k< k_l$, and within this range the wavenumbers $k$ are assumed to scale with $k_l$. Upon integrating over this range, the local integral in (4.20) yields,

\begin{eqnarray}
  K^l (p)  &\sim&  {2k_l\over 3-p} F_l \varepsilon^{1/3} 
                           L^{(5/3-p)/2} \left({ 1-\left({k_*\over k_l }\right)^{(3-p)/2} }\right) \sigma_l^2	
\end{eqnarray}

Let the size of an local eddy in wavenumber space be defined by, $C(p,R_l)=k_l/k_*$, where $C$ is finite and greater than unity. $R_l=k_l/k_1$, is related to a local Reynolds number through, $Re_l\sim R_l^{4/3}$. Thus, all dependencies on $R_l$ can be replaced by dependencies on $Re_l$, e.g. $C=C(p,Re_l)$. We will use $R_l$ in the current analysis. Then (4.21) becomes,

\begin{eqnarray}
  K^l (p)  &\sim& {2 F_l\over 3-p} \varepsilon^{1/3} L^{(5/3-p)/2}
                          k_l^{(3-p)/2} \left({1-\left({1\over C}\right)^{(3-p)/2} }\right) \sigma_l^2.	
\end{eqnarray}

Using the scaling $k_l\sim 1/\sigma_l$, this becomes,

\begin{eqnarray}
	K^l (p)  &\sim& F_l \varepsilon^{1/3} L^{(5/3-p)/2} \sigma_l^{\gamma_p^l }, 
     \qquad{\rm where}\quad   \gamma_p^l = (1+p)/2
\end{eqnarray}

Equation (4.23) reproduces the correct generalized locality scaling, $\gamma_p^l=(1+p)/2$, \cite{Morel1974}, \cite{Malik1996}. For Kolmogorov turbulence, $p=5/3$, this gives, $K_{kol}^l\sim \sigma_l^{4/3}$, which recovers the Richardson's 4/3-scaling law. This validates the mathematical formulation proposed here, justifying the various closures and scaling assumptions that have been made.

\subsection{Non-locality: the influence of non-local scales}

The important step now is the natural inclusion of the non-local contribution, $K^{nl}$. As we have observed, {\em a priori} there is no reason to neglect this term in equation (4.20) -- it is locality that has actually emerged in the present approach as an ad hoc assumption. 

The non-local contribution is the first term on the right hand side in equation (4.20),

\begin{eqnarray}
	K^{nl}(p) &\sim& \varepsilon^{1/3} L^{(5/3-p)/2} 
                        \left({ \int_{nl}k^{(1-p)/2} dk }\right) \sigma_l^2	
\end{eqnarray}

This is equivalent to strained relative motion; each scale of turbulence that is non-local to the pair separation in the wavenumber range, $k_1<k<k_*$, will set up a small straining field in the neighbourhood of the pair separation which will alter the rate of  increase of the pair separation. Previous theories have always assumed that such non-local effects are negligible. However, there are three factors that suggest that this may be an oversimplification.

Firstly, the non-local wavenumbers, $k_1<k<k_*$, possess much greater energies than at the local separation wavenumber $k_l$, and this will increase their relative influence in the pair diffusion process. 

Secondly, the time scale of the non-local scales, $T^{nl} (k)$, are much larger than the local turnover time scale, $T^{nl} (k)\gg T_l\sim 1/\omega_l$. This means that the straining fields set up by non-local wavenumber will persist for longer times than the local eddy time scale $T_l$. This will also enhance their effectiveness in the pair diffusion process. 

Thirdly, although an individual non-local wavenumber may indeed have a weak influence, the integral is over a large part of the energy spectrum. Again, this will enhance the effectiveness of the non-local scales in the pair diffusion process.

Taken all together, there is a fair chance that the total non-local contributions are significant. 
Changing variables in the integrand in equation (4.24) to $s=k/k_1$, yields

\begin{eqnarray}
	K^{nl} (p) &\sim& \varepsilon^{1/3} L^{(5/3-p)/2} 
       k_1^{(3-p)/2} \left({ \int_{nl}〖s^{(1-p)/2} ds}\right) \sigma_l^2	
\end{eqnarray}

Inside the integral in equation (4.25) the assumption of non-locality implies that the wavenumbers in range of integration, $(k_1,k_*)$, do not scale with $k_l$; thus the integral inside the brackets is just a definite integral producing a non-dimensional number, $S_{nl}=S_{nl} (p,R_l,C)$;  $S_{nl}$ is not expected to be a universal constant. This gives,

\begin{eqnarray}
	K^{nl} (p) &\sim& S_{nl} \varepsilon^{1/3} L^{(5/3-p)/2} k_1^{(3-p)/2} \sigma_l^2
\end{eqnarray}

If the upper end of the inertial subrange is assumed to scale with the large scale, $k_1\sim 1/L$, then this simplifies to, 

\begin{eqnarray}
	K^{nl} (p) &\sim& S_{nl} \varepsilon^{1/3} L^{-2/3} \sigma_l^{\gamma_p^{nl}},    
     \qquad {\rm with} \quad  \gamma_p^{nl}=2, \quad    1<p\le 3
\end{eqnarray}

$\gamma_p^{nl}$ is the non-locality scaling, and it is equal to $2$, independent of $p$. $K^{nl}$ is thus always strain dominated being proportional to $\sigma_l^2$.

\subsection{A general expression for the pair diffusivity}

The overall scaling for the turbulent pair diffusivity is the sum of the local and non-local contributions, viz

\begin{eqnarray}
   K(p,\sigma_l ) &\sim & O\left({ F_l \varepsilon^{1/3} L^{(5/3-p)/2} \sigma_l^{\gamma_p^l }  }\right)+
                                     O\left({ S_{nl} \varepsilon^{1/3} L^{-2/3} \sigma_l^2  }\right), 
                                    \quad     1<p\le 3, 	
\end{eqnarray}

or simply,

\begin{eqnarray}
	K(p) &\sim& O\left({ \sigma_l^{\gamma_p^l} }\right)+O\left({ \sigma_l^{\gamma_p^{nl}} }\right),  \qquad 1<p\le 3,  
\end{eqnarray}

where $\gamma_p^l=(1+p)/2$ is the locality scaling, and $\gamma_p^{nl}=2$ is the non-locality scaling.

Whereas previous theories for pair diffusion have been based upon simple scaling rules, here a more detailed mathematical approach has been developed by expressing the pair diffusivity through a Fourier integral decomposition. {\em a priori} assumptions regarding locality have {\em not} been made, and this has led to  an expression for $K$ as the sum of local and non-local contributions in equation (4.29). A feature of this approach is that it uncovers various scalings and closure assumptions that are essential in {\em any theory} for pair diffusion. Such assumptions are inherent within theories based upon locality scaling rules, but they are often unstated.This is clear from section 4.6 where the correct locality limit was derived using same set of assumptions and closures that has led to the more general expression in equation (4.29).

\subsection{The balance of local and non-local processes}

The critical question is, what this implies for the overall scaling in $K$, and how does the diffusivity $K$ actually manifest as a function of  $\sigma_l$?  It is clear from (4.29) that both local and non-local processes contribute to the pair diffusion process in the range, $1<p\le 3$, but there are many uncertainties in their respective magnitudes, and it is not apparent what the relative balance of their contributions is for any given, $p$, and at any given separation $\sigma_l$.

Uncertainty comes from a number of sources, most importantly from the inherent uncertainty in the size of an eddy in wavenumber space, $C(p,R_l)$, of which only a few broad properties can be deduced. It is finite, $1<C<\infty$, and depends upon $p$ and $R_l$, and upon the energy spectrum $E(k)\sim k^{-p}$.  $\log (E(k))$ becomes shallower as $p\to 1$, so there is relatively more energy in the local scales as $p$ approaches $1$, which indicates that $C$ must increase in this limit, $C\gg 1$. On the other hand, as $p\to 3$, then $C\to 1$. 

There is also uncertainty in the unknown pre-factors $F_l$ and $S_{nl}$ which account for the various scalings and closures assumed in the analysis.

It would seem impossible to go beyond this point since almost nothing else is known about the behaviour of  $C$. Fortunately, this problem can be approached from another angle in which neither the exact form of $C$ nor the exact balance of the local and non-local contributions need to be calculated in order to deduce the general scaling for $K$. It is sufficient to obtain the asymptotic behaviour of $C$ at $p=1$ and $p=3$, and the general trend and continuity in the balance as a function of $p$ in the range $1<p<3$. These are obtained as follows. 

The balance of the local and non-local diffusivities is defined as the ratio, $M_K (p,R_l,C) = K^{nl}/K^l$. $M_K$ is a function of, $p, R_l$, and $C$, and it is calculated using equations (4.22), (4.23), (4.25), and (4.26), leading to

\begin{eqnarray}
	M_K &=& \displaystyle{K^{nl}\over K^l} \sim 
                    {\left({1-\left(\displaystyle{C\over R_l}\right)^{(3-p)/2)}}\right) \over 
                F_l \left({C^{(3-p)/2)}-1}\right)  }
\end{eqnarray}

As, $p\to 1$, then $M_K\to (1-C/R_l)/F_l (C-1)$. For a large inertial subrange, $R\gg 1$, and with $F_l\approx 0.5$ and $1\ll C\ll R_l$, then $M_K\to 1/F_l R_l\ll 1$, and therefore locality dominates in this limit. 

As, $p\to 3$, then $M_K\to \ln⁡ R_l/\ln ⁡C-1$; and with $F_l\approx 0.5$ and, $1<C\ll R_l^{2/3}$, then $M_K\gg 1$, and therefore non-locality dominates in this limit. 

In the intermediate range, $1<p<3$, although the balance $M_K$ cannot be obtained quantitatively without the explicit form of  $F_l$ and $C$ as a functions of $p$ and $R_l$, it is instructive to investigate $M_K$ with a wide range of test functions for $C$ to see if any essential trends in the behaviour of $M_K$ can be elucidated. For this purpose, some simplifications are assumed in the following. It is assumed that, $F_l=0.5$, and for separations deep inside the inertial subrange we set, $R_l=10^4$. $C$ is then a function of $p$ in these test cases. 

The first set of test functions for $C$ are exponential functions, $C(p)=1.1+ (c_1-1.1)\exp⁡(-5(p-1))$, with $C(1)=c_1$, and $C(3)\approx 1.1$. The second set of test functions for $C$ are linear functions, $C(p)=c_1-(c_1-1.1)(p-1)/2$, with $C(1)=c_1$, and $C(3)=1.1$. A third set of test functions for $C$ are constants, $C(p)=c_1$. In each of these three sets of test cases, $M_K$ was calculated using (4.30) for five selected values of $c_1=4,10,20,50$, and $100$.

For all the test functions considered, $M_K\gg 1$ as $p\to 3$, (Figs. 2, 3 and 4), indicating that non-locality is always dominant in this limit. Furthermore, $M_K\ll 1$ as $p\to 1$, in all the test cases considered, indicating that locality is always dominant in this limit. 

Importantly, $M_K (p)$ increases smoothly and monotonically as $p$ increases from $1$ to $3$ in all the test cases considered. The linear and constant test functions display very similar trends (Figs. 3 and 4), while the exponential test functions (Fig. 2), differ from these by up to a factor of about 10 at any value of $p$.

Thus, the asymptotic limits of $M_K$ at $p=1$ and $p=3$ are observed to be only weakly dependent upon the form of $C$,  and $M_K$ is observed to be a smooth function of $p$ in the range $1<p<3$. 
The following conclusions may therefore be drawn for large inertial subranges where $R_l\gg 1$. Firstly, as $p\to 1$ then $M_K\ll 1$, and therefore $K^{nl}\ll K^l$,  yielding the locality limit, 

\begin{eqnarray}
	K(p) &\to& K^l (1)  \sim \sigma_l^1 \qquad {\rm as} \quad p\to 1.
\end{eqnarray}

\begin{figure}
\begin{center}
\mbox{\subfigure{\includegraphics[width=10cm]{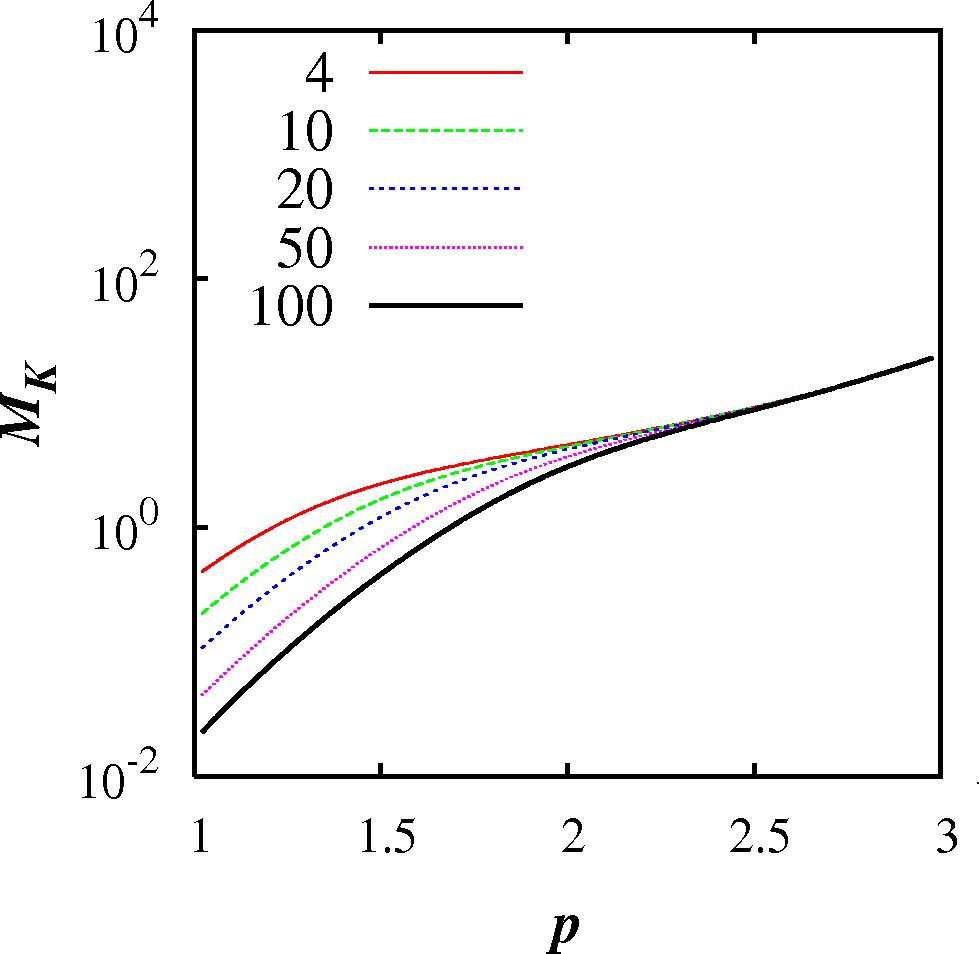}}}
{\linespread{1.2}
\caption{The balance $M_K=K^{nl}/K^l$ shown as $\log(M_K)$ against $p$. Exponential test functions, $C(p)=1.1+ (c_1-1.1) \exp⁡(-5(p-1))$, were used in $M_K$ in equation (4.30) with $R_l=10^4$, and $F_l=0.5$. Five cases are shown, $c_1=4,10,20,50,100$. \protect}\label{fig02}
}
\end{center}
\end{figure}

\begin{figure}
\begin{center}
\mbox{\subfigure{\includegraphics[width=10cm]{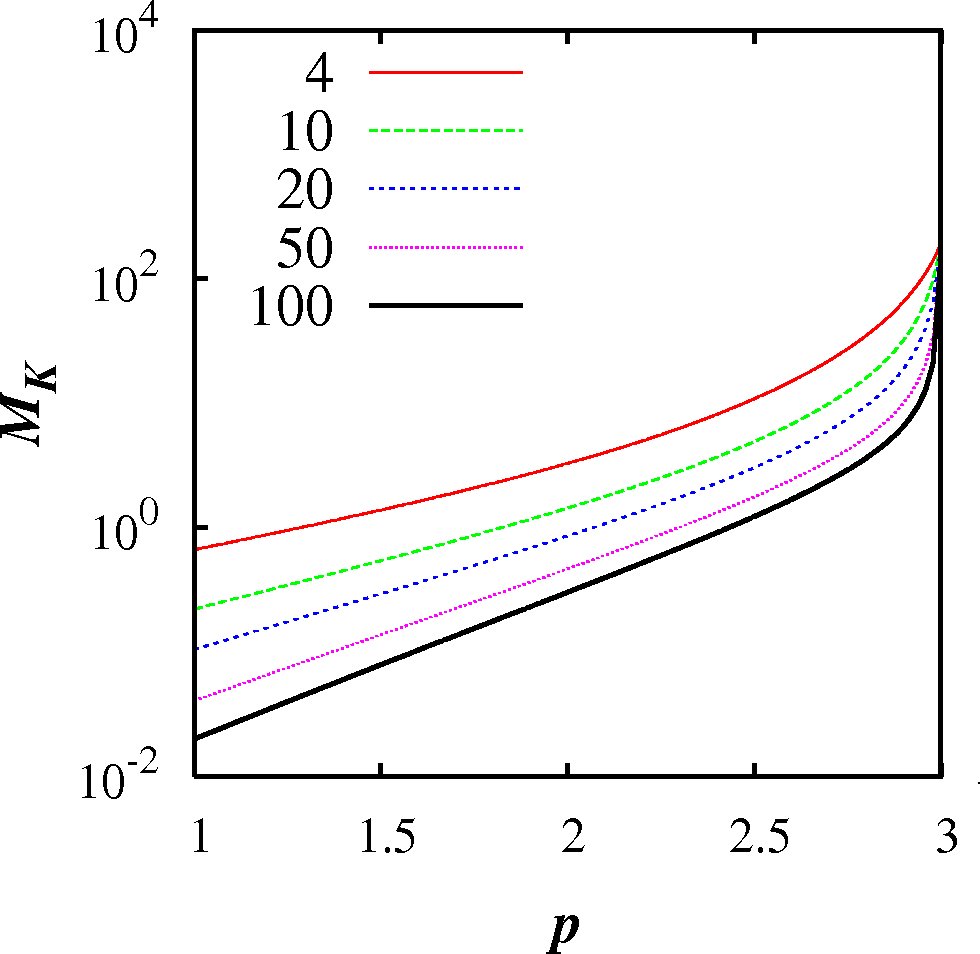}}}

{\linespread{1.2}
\caption{The balance $M_K=K^{nl}/K^l$ shown as $\log(M_K)$ against $p$. Linear test functions, $C(p)=c_1-(c_1-1.1)(p-1)/2$, were used in $M_K$ in equation (4.30) with $R_l=10^4$, and $F_l=0.5$. Five cases are shown, $c_1=4,10,20,50,100$.          \protect}\label{fig03}
}
\end{center}
\end{figure}

\begin{figure}
\begin{center}
\mbox{\subfigure{\includegraphics[width=10cm]{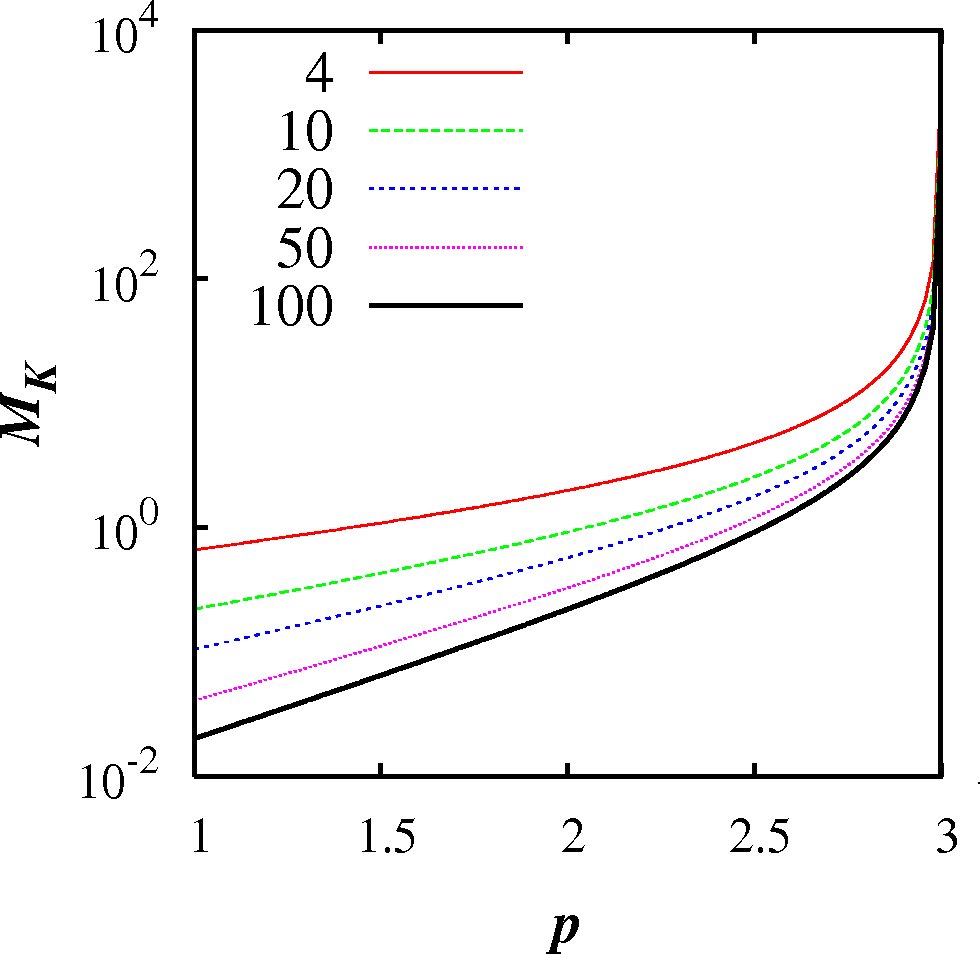}}}

{\linespread{1.2}
\caption{The balance $M_K=K^{nl}/K^l$ shown as $\log(M_K)$ against $p$. Constant test functions, $C(p)=c_1$, were used in $M_K$ in equation (4.30) with $R_l=10^4$, and $F_l=0.5$. Five cases are shown, $c_1=4,10,20,50,100$.           \protect}\label{fig04}
}
\end{center}
\end{figure}

Secondly, as $p\to 3$ then $M_K\gg 1$, and therefore $K^{nl}\gg K^l$,  yielding the non-locality limit,

\begin{eqnarray}
	K(p) &\to& K^{nl} (3)  \sim \sigma_l^2  \qquad {\rm as} \quad p\to 3.
\end{eqnarray}

Thirdly, $M_K (p)$ must also be a smooth function of $p$ in the range $1<p\le 3$  and it increases smoothly and monotonically as $p$ increases smoothly from $1$ to $3$; the non-local scales exert increasingly stronger influence at the local pair separation scale until they are completely dominant.

\subsection{The general scaling for $K$}

Since $M_K (p)$ is a smooth and continuous function of $p$ in the range $1<p\le 3$, then $K(p)$ must also be a smooth and continuous function of $p$ in this range and it must display a smooth transition between its asymptotic limits, $K(1) \sim\sigma_l^1$ and $K(3)\sim  \sigma_l^2$, as $p$ passes smoothly from $1$ to $3$.

It is possible that either one of the local or non-local processes dominates throughout the inertial subrange for any given $p$; but that would imply a discontinuous jump between locality and non-locality scalings at some value of $p$ in order to satisfy the asymptotic limiting cases in (4.31) and (4.32). This is unlikely, and also violates the continuity in $K(p)$ as a function of $p$.

It is also possible that for every $p$ both local and non-local power laws exist in different parts of the inertial subrange, with a crossover between them in the middle of the inertial subrange. In this case some new, as yet unknown, crossover separation scale $l_c (p)$ has to be introduced in to the theory. This is also unlikely, and no evidence for such a cross-over length scale exists.

A third possibility is that both local and non-local diffusional processes are effective in the pair diffusion process inside the inertial subrange for all $p$ in the range, $1<p\le 3$, at all separations where the new theory is valid. Then, for any given turbulence spectrum, $E(k)\sim k^{-p}$, the pair diffusivity $K(p)$ must manifest as a smooth transition between the two asymptotic limits, (4.31) and (4.32), over the range of separations for which it is valid. Since $K(p)$ also satisfies (4.29) then it must be a power law scaling which is intermediate between the purely local and non-local scalings,

\begin{eqnarray}
	K(p) &\sim& \sigma_l^{\gamma_p}, \qquad {\rm with}\quad  
             \gamma_p^l < \gamma_p< \gamma_p^{nl}, \qquad {\rm for} \quad   1<p< 3.
\end{eqnarray}

The new scaling powers $\gamma_p$ are such that, as $p\to 1$ then  $\gamma_p\to 1$, and as $p\to 3$ then $\gamma_p\to 2$. Globally, $1<\gamma_p\le 2$. Furthermore, $\gamma_p$ must transform smoothly between these limiting cases as $p$ goes from $1$ to $3$. Note that $\gamma_p=2$ is attained only when, $p=3$.

Equation (4.33) is the main prediction of the new theory, with  $\gamma_p^l=(1+p)/2$ and  $\gamma_p^{nl}=2$. It is equivalent to the mean square separation scaling, $\langle l^2\rangle  \sim t^{\chi_p}$, with $\chi_p$ given by $\chi_p=1/(1-\gamma_p/2)$ from equation (4.1). 

For Kolmogorov turbulence, $E\sim k^{-5/3}$, the new theory predicts that, $\gamma_{Kol}>4/3$. For real turbulence which includes intermittency $\mu_I>0$, such that $E\sim k^{-(5/3+\mu_I)}$, the scaling should be higher with $\gamma_{\mu_I} > \gamma_{\mu_I}^l=4/3+\mu_I/2$. 

The new theory implies that a non-Richardson 4/3-law, $K\sim \sigma_l^{4/3}$, exists for some spectrum, $E\sim k^{-p_*}$,  with $p_*< 5/3$ where $\gamma_{p_*}=4/3$. This is equivalent to $\langle l^2\rangle  \sim t^3$, which is a new $t^3$-regime for the mean square separation.

The ratio of the non-local and local power scaling is defined as,

\begin{eqnarray}
	M_\gamma (p) &=& \displaystyle{\gamma_p\over \gamma_p^l}
\end{eqnarray}

$M_\gamma (p)$ is equal to $1$ at both $p=1$ and $p=3$, and since $M_\gamma >1$  in the range $1<p<3$, then there must be a maximum in $M_\gamma$ at some $p=p_m$ for an energy spectrum $E\sim k^{-p_m}$, where $1<p_m<3$.

The new theory predicts only the upper and lower bounds for the scaling powers, equation (4.33), but it holds for a wide range of power spectra, $1<p\le 3$. The available data exists only for a single spectrum, although a very important one, for real turbulence including intermittency for which the geophysical data, Fig. 1, suggests a scaling of, $\gamma_{\mu_I}\approx 1.564$, i.e. $K_{\mu_I}(p)\sim \sigma_l^{1.564}$.

\section{Numerical calculations using a Lagrangian diffusion model}

We have developed a new non-local theory for turbulent particle pair diffusion, and seen that it predicts scaling laws for the pair diffusivity, $K$, equation (4.33), which is consistent with the reappraised 1926 data from geophysical turbulence, Fig. 1. 

However, the values of $\gamma_p$ cannot be predicted from the theory alone. To do this, and to examine other predictions of the new theory, ideally we need experiments and/or DNS. Unfortunately, current capabilities of both experiments and DNS are far from being able to examine very large inertial subrange turbulence, as seen in Sections 2 and 3. However, we can make some progress in examining the theory through the use of Lagrangian diffusion models.

Lagrangian diffusion models, like random walk models, do not necessarily satisfy Navier-Stokes equations, but they can be indicative of the scaling behaviour of Lagrangian statistics in fluid flows, sometimes quite accurately. Furthermore, neither is Richardson's theory based upon satisfying Navier-Stokes equation, so in the present study Lagrangian diffusion models may be able to shed some useful light on to this problem.

\subsection{Particle trajectories using KS}

In this study, the Lagrangian diffusion model Kinematic Simulations (KS) was used to obtain the statistics of particle pair diffusion. In KS one specifies the second order Eulerian structure function through the power spectrum, \cite{Kraichnan1970}, \cite{Fung1992}. KS is particularly useful here because it can generate extended inertial subranges sufficient to test inertial subrange pair diffusion scaling laws. Pair diffusion statistics were thus obtained from KS containing generalized energy spectra, $E(k)\sim k^{-p}$, $k_1\le k\le k_\eta$ over extended inertial subranges with $k_\eta/k_1=10^6$, and for $1<p\le 3$. 

KS generates turbulent-like non-Markovian particle trajectories by releasing particles in flow fields which are prescribed as sums of energy-weighted random Fourier modes. By construction, the velocity fields are incompressible and the energy is distributed among the different modes by a prescribed Eulerian energy spectrum, $E(k)$. The essential idea behind KS is that the flow structures in it  - eddying, straining, and streaming zones - are similar to those observed in turbulent flows, although not precisely the same. This is sufficient to generate turbulent-like particle trajectories.

KS has been used to examine single particle diffusion \cite{Turfus1987}, and pair diffusion \cite{Fung1992}, \cite{Malik1996}, \cite{Fung1998}, \cite{Malik1999}, \cite{Nicolleau2011}.  KS has also been used in studies of turbulent diffusion of inertial particles \cite{Maxey1987}, \cite{Meneguz2011}, \cite{Farhan2015}. Meneguz \& Reeks found that the statistics of the inertial particle segregation in KS generated flow fields for statistically homogeneous isotropic flow fields are similar to those generated by DNS. KS has also been used as a sub-grid scale model for small scale turbulence \cite{Perkins1993}, \cite{Yao2009}.

\subsection{The KS velocity fields and energy spectra}

An individual Eulerian turbulent flow field realization is generated as a truncated Fourier series,

\begin{eqnarray}
   {\bf u}({\bf x},t) &=& 
   \sum_{n=1}^{N_k}\left({({\bf A}_n\times {\bf \hat k}_n )\cos⁡({\bf k}_n\cdot {\bf x}+\omega_n t) +
                                         ({\bf B}_n\times {\bf \hat k}_n )\sin⁡({\bf k}_n\cdot {\bf x}+\omega_n t) }\right) 	
\end{eqnarray}

 where $N_k$ is the number of representative wavenumbers, typically hundreds for very long spectral ranges, $k_\eta/k_1\gg 1$. $\hat k_n$ is a random unit vector; ${\bf k}_n=k_n \hat k_n$ and $k_n=|{\bf k}_n |$. The coefficients ${\bf A}_n$ and ${\bf B}_n$ are chosen such that their orientations are randomly distributed in space and uncorrelated with any other Fourier coefficient or wavenumber, and their amplitudes are determined by $\langle {\bf A}_n^2 \rangle= \langle {\bf B}_n^2  \rangle\propto E(k_n )dk_n$, where $E(k)$ is the energy spectrum in some wavenumber range $k_1\le k\le k_\eta$. The angled brackets $\langle\cdot\rangle$ denotes the ensemble average over particles and over many random flow fields. The associated frequencies are proportional to the eddy-turnover frequencies, $\omega_n=\lambda\sqrt{k_n^3 E(k_n)}$.  There is some freedom in the choice of $\lambda$, so long as $0\le \lambda<1$. The construction in equation (5.1) ensures that the Fourier coefficients are normal to their wavevector which automatically ensures incompressibility of each flow realization, $\nabla\cdot u=0$. The flow field ensemble generated in this manner is statistically homogeneous, isotropic, and stationary.

Unlike some other Lagrangian methods, by generating entire kinematic flow fields in which particles are tracked KS does not suffer from the crossing-trajectories error which is caused when two fluid particles occupy the same location at the same time in violation of incompressibility; but because KS flow fields are incompressible by construction this error is completely eliminated.

Intermittency exists in KS because flow structures such as vortex tubes exist in the flow field, although they are not exactly the same as in fully dynamic turbulence, \cite{Fung1992}. By adjusting the power spectrum we can mimic the intermittent spectrum seen in real turbulence. KS pair diffusion statistics have been validated and found to produce close agreement with DNS, at least for low Reynolds numbers, \cite{Malik1999}, incuding the flatness factor of pair separation.

The energy spectrum $E(k)$ can be chosen freely within a finite range of scales, even a piecewise continuous spectrum, or an isolated single mode are possible. To incorporate the effect of large scale sweeping of the inertial scales by the energy containing scales, the simulations are carried out in the sweeping frame of reference by setting $E(k)=0$ in the largest scales, for $k<k_1$, and an inverse power spectrum in the inertial subrange,

\begin{eqnarray}
	E(k)=C_k ε^{2/3} L^{5/3-p} k^{-p}, \qquad k_1\le k\le k_\eta,\quad  1<p\le 3
\end{eqnarray}

where $C_k$ is a constant. The largest represented scale of turbulence is $2\pi/k_1$ and smallest is the Kolmogorov micro-scale $\eta=2\pi/k_\eta$. $L$ is some large length scale, such as the integral length scale, or a Taylor length scale.

A particle trajectory, ${\bf x}_L (t)$, is obtained by integrating the Lagrangian velocity, ${\bf u}_L (t)$, in time,

\begin{eqnarray}
	{d{\bf x}_L \over dt} &=& {\bf u}_L (t)={\bf u}({\bf x},t).
\end{eqnarray}

Pairs of trajectories are harvested from a large ensemble of flow realizations and pair statistics are then obtained from it for analysis.

\subsection{KS Parameters}

The spectrum in (5.2) is normalized such that the total energy density contained in any flow realization is $3u'/2$, where $u'$ is the rms turbulent velocity fluctuations in each direction.  

In the current simulations, $k_1=1, L=1,  C_k=1.5$ (Kolmogorov constant) and  $u'=1$. Then this yields,

\begin{eqnarray}
	\varepsilon^{2/3} &=& (p-1)\left({ 1-\left({ k_1\over k_\eta }\right)^{p-1} }\right)^{-1}	
\end{eqnarray}

$p=1$ is a singular limit which is not consider here. The size of the inertial subrange is fixed in all the simulations here at, $k_\eta/k_1=10^6$.  With (5.4), $v_\eta=(\varepsilon \eta)^{1/3}$  is the velocity micro-scale, and $t_\eta=\varepsilon^{-1/3} \eta^{2/3}$ is the time micro-scale.

The distribution of the wavenumbers is geometric, $k_n=r^{n-1} k_1$, and the common ratio is $r=(k_\eta/k_1 )^{1/(N_k-1)}$. The increment in wavenumber-space of the n’th wavenumber is $\Delta k_n=k_n (\sqrt r-1/\sqrt r)$.  The frequency factor is chosen to be, $\lambda=0.5$, which is typical in many KS studies. A choice of $\lambda<1$ does not affect the scaling in the pair diffusivity, even frozen turbulence, $\lambda=0$, has been found to yield the same scaling, which has been attributed to the open streamline topology of streamlines in 3D flows, \cite{Malik1996}.

It is important to simulate cases over the whole range of in the energy spectrum $1<p\le 3$, in order to examine the new theory comprehensively.

\subsection{Simulation Results}

The spectra in equation (5.2) were prescribed in the KS simulations. 25 cases of $p$ were selected in the range, $1<p\le 3$. The case $p=1$ is singular, but $p$ can be taken close to this limit; the smallest value of $p$ chosen was, $1.01$ (Table 2). 

The particle trajectories were obtained by integrating (5.3) using the 4th order Adams-Bashforth predictor-corrector method (4th order Runga-Kutta gives identical results). \cite{Thomson2005} used a variable time step ∆t that was small compared to the turnover time of an eddy of the size of the instantaneous pair separation, but larger than the turbulence micro-scale $t_\eta$. While this speeds up the turnaround time of the calculations, here it is desired to avoid extra assumptions so that unambiguous conclusions can be drawn from the results. Therefore, in all of the current simulations a very small but fixed time step, $\Delta t\ll t_\eta$ has been used. This has the further advantage of avoiding any smoothening of the particle trajectories that is necessary when using variable time steps.

Eight pairs of a particles were released in each flow realization, placed initially at the corners of a cube of side $3$ units, beginning at $(0,0,0)$. This is far enough apart for each pair to be independent. It is crucial to run over a large number of different flow realizations, otherwise the statistics will not converge. Typically the ensemble was around $5000$ flow realizations, yielding a harvest of $40,000$ particle pair trajectories per simulation. A simulation was run for about one large timescale, $T=2\pi/k_1$, which required around $10^6$ time steps.

\begin{table}
\begin{center}
\def~{\hphantom{0}}
  \begin{tabular}{lll lll}

$p$	&$\gamma_p$	&$\gamma_p^l$	&$\chi_p$	&$\chi_p^l$	&$M_\gamma$ \\
&&&&& \\
1.01	&1.060	&1.005	&2.128	&2.010	&1.055\\[-3pt]
1.1		&1.120	&1.05	&2.273	&2.105	&1.067\\[-3pt]
1.2		&1.190	&1.10	&2.469	&2.222	&1.082\\[-3pt]
1.3		&1.260	&1.15	&2.703	&2.353	&1.096\\[-3pt]
1.4		&1.340	&1.20	&3.030	&2.500	&1.117\\[-3pt]
1.5		&1.410	&1.25	&3.390	&2.667	&1.128\\[-3pt]
1.6		&1.480	&1.30	&3.846	&2.857	&1.139\\[-3pt]
5/3		&1.525	&4/3	&4.211	&3		&1.144\\[-3pt]
1.70	&1.545	&1.35	&4.396	&3.077	&1.144\\[-3pt]
1.72	&1.555	&1.360	&4.494	&3.125	&1.143\\[-3pt]
1.74	&1.570	&1.370	&4.651	&3.175	&1.146\\[-3pt]
1.77	&1.585	&1.385	&4.819	&3.252	&1.144\\[-3pt]
1.8		&1.605	&1.40	&5.063	&3.333	&1.146\\[-3pt]
1.9		&1.660	&1.45	&5.882	&3.636	&1.145\\[-3pt]
2.0		&1.710	&1.50	&6.897	&4		&1.140\\[-3pt]
2.1		&1.750	&1.55	&8.000	&4.444	&1.129\\[-3pt]
2.2		&1.790	&1.60	&9.424	&5		&1.119\\[-3pt]
2.3		&1.820	&1.65	&11.11	&5.714	&1.103\\[-3pt]
2.4		&1.850	&1.70	&13.33	&6.676	&1.088\\[-3pt]
2.5		&1.880	&1.75	&16.67	&8		&1.074\\[-3pt]
2.6		&1.900	&1.80	&20.00	&10		&1.056\\[-3pt]
2.7		&1.930	&1.85	&28.57	&13.33	&1.043\\[-3pt]
2.8		&1.950	&1.90	&40.00	&20		&1.026\\[-3pt]
2.9		&1.970	&1.95	&67.67	&40		&1.010\\[-3pt]
3.0		&2.000	&2.00	&$\infty$&$\infty$&1.000\\[3pt]
 \end{tabular}

{\linespread{1.2}
 \caption{
$p, \gamma_p,\gamma_p^l, \chi_p, \chi_p^l$, and $M_\gamma$. The pair diffusivity is, $K\sim \sigma_l^{\gamma_p}$, the locality scaling is, $\gamma_p^l=(1+p)/2$. The mean square separation is, $\langle l^2\rangle=t^{\chi_p}$, where $\chi_p=1/(1-\gamma_p/2)$, and the locality scaling is, $\chi_p^l=1/(1-\gamma_p^l/2)$. The ratio of the power scalings is, $M_\gamma =\gamma_p/\gamma_p^l$.\protect}
}
 \end{center}
 \end{table}

\begin{figure}
\begin{center}
\mbox{\subfigure{\includegraphics[width=10cm]{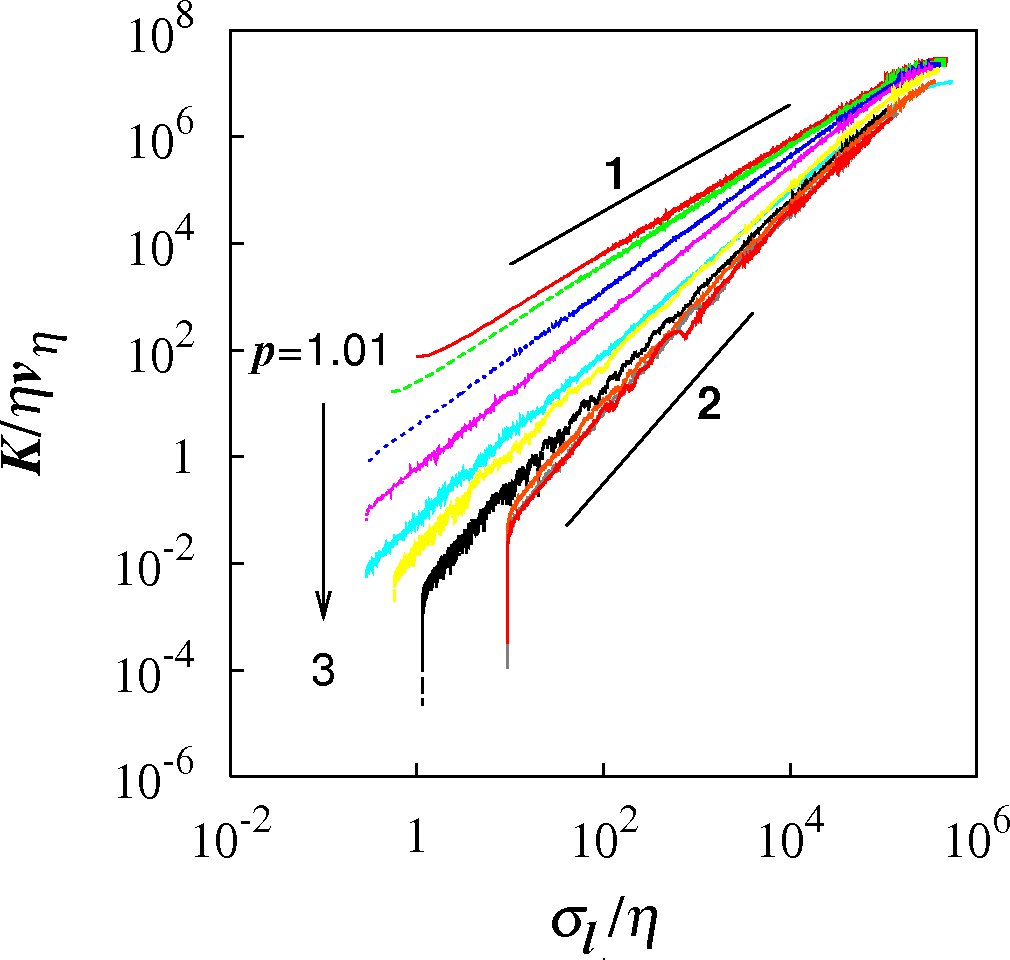}}}

{\linespread{1.2}
\caption{
The turbulent pair diffusivity as $\log(K/\eta v_\eta)$ against $\log(\sigma_l/\eta)$. The results were obtained from Kinematic Simulations. The KS energy spectra used were, $E(k)\sim k^{-p}$, with the inertial subrange, $k_\eta/k_1=10^6$.  25 cases cover the range $1<p\le 3$, (Table 2); but for clarity only 10 cases are shown here: $p=1.01,1.1,1.3,1.5,5/3,1.9,2.2,2.5,2.9,3.0$. The slopes, $\gamma_p$, show a smooth transition from $p=1.01$ to $3$. Solid black lines of slopes 1 and 2 are shown for comparison.\protect}\label{fig05}
}
\end{center}
\end{figure}

\begin{figure}
\begin{center}
\mbox{\subfigure{\includegraphics[width=10cm]{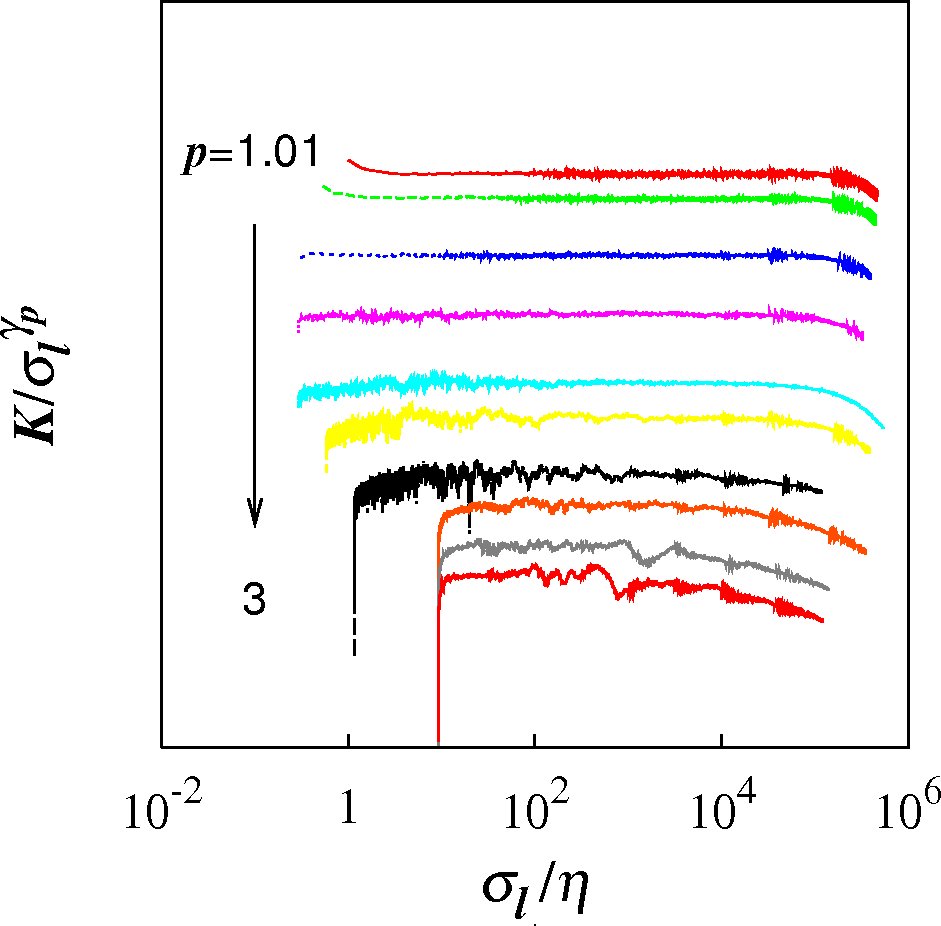}}}

{\linespread{1.2}
\caption{
Compensated turbulent pair diffusivity as $\log(K/\sigma_l^{\gamma_p})$ against $\log(\sigma_l/\eta)$. The colour coding is similar to that in Figure 5. The values of, $\gamma_p$, (Table 2) are the ones that display the best near horizontal line over the longest separation range. For clarity some of the plots have been spaced out vertically, hence no scale is shown on the ordinate. This does not affect the scaling which is the main interest here.\protect}\label{fig06}
}
\end{center}
\end{figure}

The results obtained from KS are summarized in Table 2. This table shows $\gamma_p$ from the simulations in column2; $\gamma_p^l=(1+p)/2$ in cloumn 3. The equivalent mean square separation scaling laws, $\langle l^2 \rangle \sim t^\chi$, are also shown: $\chi_p=1/(1-\gamma_p/2)$  is the non-local scaling, column 4; $\chi_p^l=1/(1-\gamma^l_p/2)=4/(3-p)$   is the local scaling, column 5.  The ratio of scaling powers, $M_\gamma (p)=\gamma_p/\gamma_p^l$ is shown in column 6.

Log-log plots of the turbulent pair diffusivity $K/\eta v_\eta$ against the separation $\sigma_l/\eta$ for different energy spectra $p$ display clear power law scalings of the form, 

\begin{eqnarray}
	K(p) &\sim& \sigma^{\gamma_p}, \qquad      1<\gamma_p\le 2, \quad   1<p\le 3,
\end{eqnarray}

over wide ranges of the separation inside the inertial subrange (Fig. 5). The ${\gamma_p}$'s are the slopes of the plots in Figure 5; these  are more easily observed by plotting the compensated relative diffusivity $K/\sigma_l^{\gamma_p}$ against the separation (Fig. 6). The  $\gamma_p$'s are the powers that give near-horizontal lines over the longest range of separation, a procedure that determines $\gamma_p$ to within $1\%$ error for most $p$, except near the singular limit,  $p=1$, where the errors are around $6\%$.

The plots of  $\gamma_p$ and $\gamma_p^l=(1+p)/2$ against $p$  show that the new scaling powers, $\gamma_p$, are in the range, $(1+p)/2<\gamma_p\le 2$, (Fig. 7, black filled circles and dotted blue line respectively, left hand scale). It is observed that as $p\to 1$, then $\gamma_p\to 1$; and as $p\to 3$, then $\gamma_p\to 2$. Furthermore, the, ${\gamma_p}$'s, display a smooth transition between the asymptotic limits at $p=1$ and $3$.

The range of separation over which the scalings, $K(p) \sim\sigma_l^{\gamma_p}$, are observed to progressively reduce from both ends of the range as $p\to 3$. It reduces from below due to the diminishing energies contained in the smallest scales so that the pair separation penetrates further into the inertial subrange before it 'forgets' the initial separation. It reduces from above because the long range correlations are increasingly stronger as $p\to 3$ causing the particles in a pair to become independent at earlier times and at smaller separation.

The plot of $M_\gamma$ against $p$ show a peak at $p_m\approx 1.8$, where $M_\gamma (p_m) \approx 1.15$, (Fig. 7, right hand scale). In fact, $M_\gamma$  remains relatively close to this peak value in a neighbourhood of $p=1.8$, in the range $1.5<p<2$.

Plots of the scaling powers, $\chi_p$, obtained from the simulations (filled black circles), and from $\chi_p^l=4/(3-p)$ (dotted blue line), against $p$ are shown in Figure 8 . The same is shown in Figure 9, but focussing in the range $1.5<p<2$ which covers the intermittent turbulence range which is indicated by the two red vertical lines. 

For Kolmogorov turbulence, $p=5/3$, KS produces, $\gamma_{Kol}\approx 1.525$, so that  $K_{Kol} \sim\sigma_l^{1.525}$.  

For real turbulence with intermittency corrections, $\mu_I > 0$, such that  $E(k)\sim k^{-(5/3+\mu_I)}$, three extra cases, $p_I=5/3+\mu_I$, where simulated. The currently accepted value for the intermittency lies in the range, $0.025<\mu_I<0.075$, \cite{Anselmet2001}, \cite{Tsuji2004}, \cite{Meyers2008}, \cite{Tsuji2009}. Therefore, three cases that cover most of this range were simulated, $p_I=1.70,1.72$,and $1.74$. For these spectra, KS produced,(Fig. 7 and Table 2),

\begin{eqnarray}
K_{\mu_I} &\sim \sigma_l^1.545,  	&\quad{\rm for}\quad p_I=1.70,\ \ \mu_I=0.033 \nonumber \\
K_{\mu_I} &\sim \sigma_l^1.555,        &\quad{\rm for}\quad p_I=1.72,\ \ \mu_I=0.053\\
K_{\mu_I} &\sim \sigma_l^1.570,        &\quad{\rm for}\quad p_I=1.74,\ \ \mu_I=0.073\nonumber
\end{eqnarray}

\begin{figure}
\begin{center}
\mbox{\subfigure{\includegraphics[width=10cm]{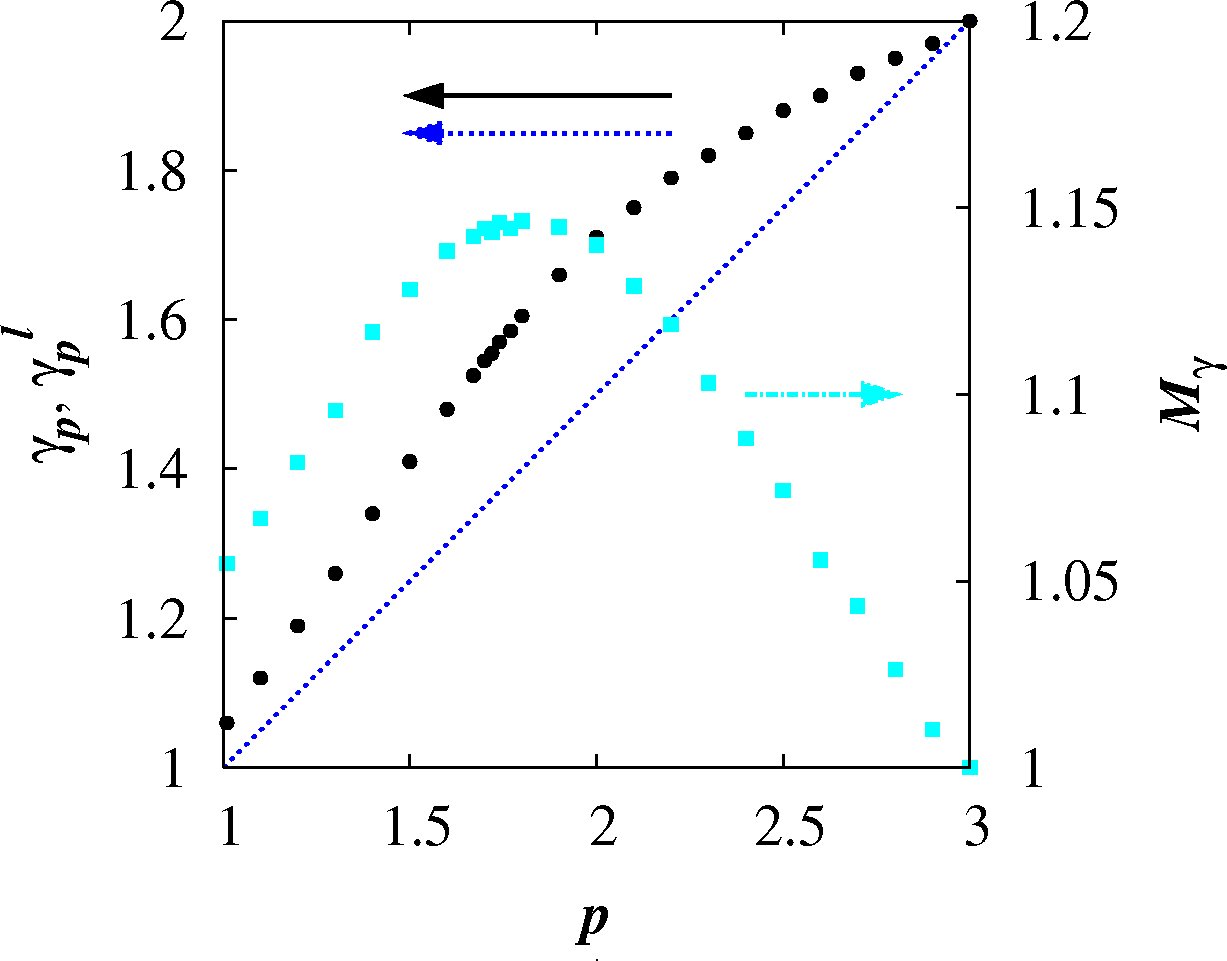}}}

{\linespread{1.2}
\caption{
$\gamma_p$, $\gamma_p^l$, and $M_\gamma$ against $p$. The black filled circles are the ${\gamma_p}$'s. The dotted blue line is the locality scaling $\gamma_p^l=(1+p)/2$. The cyan squares are the ratios, $M_\gamma=\gamma_p/\gamma_p^l$ (right hand scale), (Table 2).
\protect}\label{fig07}
}
\end{center}
\end{figure}

The mid-point in the above intermittency range is close to, $p_I=1.72$, which gives the scaling  law $K_{\mu_I} \sim \sigma_l^{1.555}$, which is an error of less than $0.6\%$ in the scaling power compared to the data in Figure 1. For, $p_I=1.70$, we obtain $K_{\mu_I} \sim \sigma_l^{1.545}$, and the error in the scaling power  is $1.2\%$; for $p_I=1.74$, we obtain $K_{\mu_I} \sim \sigma_l^{1.570}$, and the error in the scaling power is $0.4 \%$. It is reasonable to infer that the error will be almost zero close to $p_I\approx 1.73$. 

The Kolmogorov spectrum $p=5/3$ produces, $K_{Kol} \sim \sigma_l^{1.525}$, and the error in the scaling power is  $2.5\%$.  The Richardson's locality law, $K\sim \sigma_l^{4/3}$, is $15\%$ in error in the scaling power compared to the reappraised geophysical data in Figure 1. 

The simulations produce scalings for $\langle l^2\rangle\sim t^{\chi_p}$, in the range $\sim t^{4.494}$  to $\sim t^{4.615}$ for the same intermittent spectra considered above, Fig. 9, Table 2. It is interesting to note that even under the assumption of locality intermittency spectra should produce scalings in the range $\langle l^2\rangle\sim t^{3.077}$  to $\sim t^{3.252}$  -- thus, a true R-O  $t^3$-regime cannot exist in reality.

In general KS predicts the correct scaling for the turbulent pair diffusivity to within $1\%$ of the geophysical data in Figure 1 in most of the accepted intermittency range centred in the neighbourhood of, $E(k)\sim k^{-1.73}$.

Finally, KS produces a non-Richardson 4/3-scaling, $K\sim\sigma_l^{4/3}$, where  $\gamma_{p_*}=4/3$ at $p_*\approx 1.4$, (Fig. 7 and Table 1). With this spectrum, $E(k)\sim k^{-1.4}$,  this is equivalent to a new $\langle l^2\rangle\sim t^3$ regime.

\begin{figure}
\begin{center}
\mbox{\subfigure{\includegraphics[width=10cm]{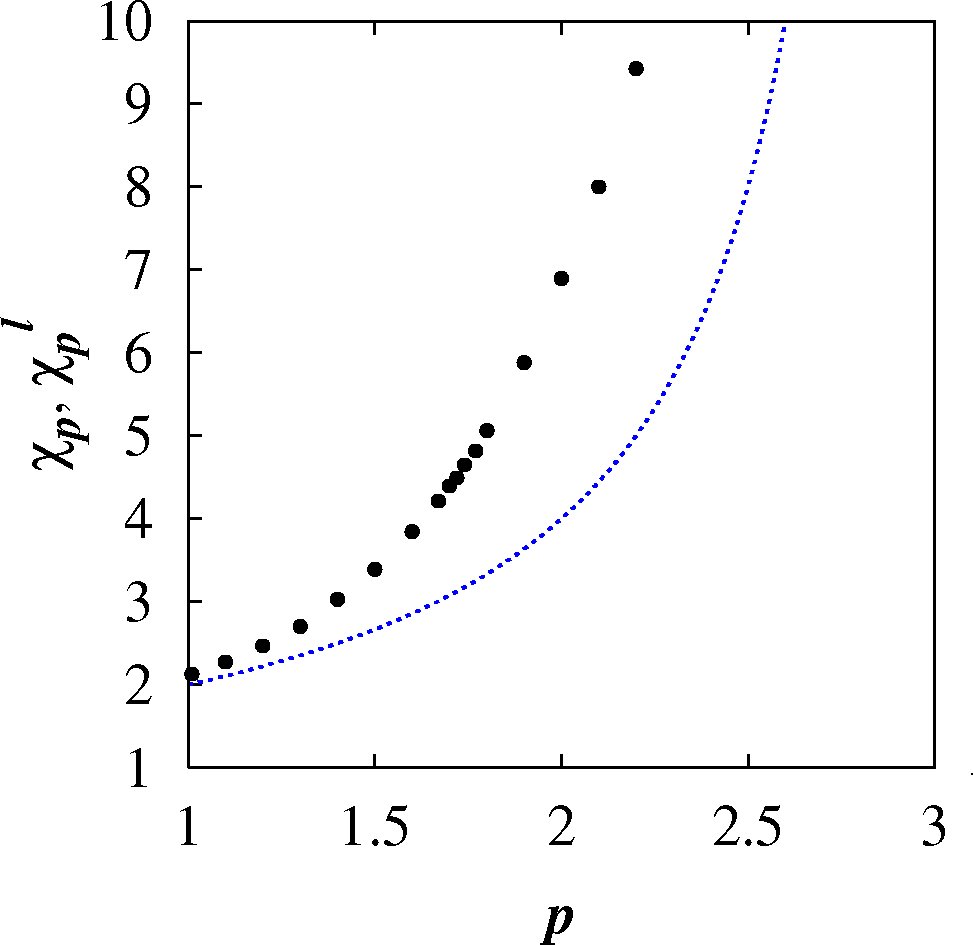}}}

{\linespread{1.2}
\caption{
$\chi_p$, $\chi_p^l$, against $p$. The black filled circles are, ${\chi_p}$'s from the simulations. The dotted blue line is the locality scaling, $\chi_p^l=4/(3-p)$.  See Table 2.\protect}\label{fig08}
}
\end{center}
\end{figure}

\begin{figure}
\begin{center}
\mbox{\subfigure{\includegraphics[width=10cm]{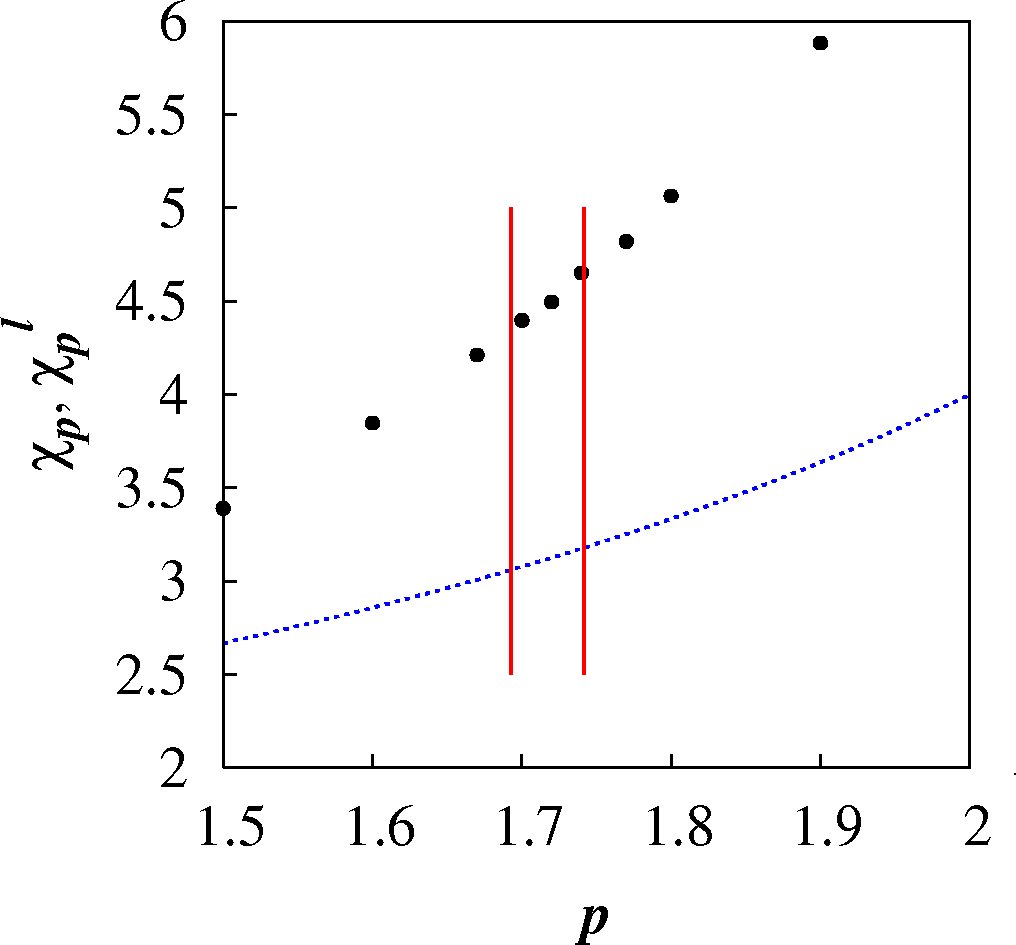}}}

{\linespread{1.2}
\caption{
$\chi_p$, $\chi_p^l$, against $p$.  Similar to figure 8, but focussing in the range of real turbulence with intermittency. The two vertical red lines are the lower and upper bounds for intermittent turbulence spectra.\protect}\label{fig09}
}
\end{center}
\end{figure}

\subsection{Estimate of numerical errors}

The numerical results presented here are the most comprehensive obtained to date from KS due to the very large ensemble of particle pairs and the small time steps used. The statistical fluctuations in the results are therefore small. The ${\gamma_p}'$s, which are the slopes of the plots in Figure 5, can be determined to within $1\%$ error. An exception is close to the singular limit $p=1$ where the numerical errors can be large. An accurate estimate of this error can be obtained as follows, noting first that the error level in $\gamma_p$ is identical to the error level in $M_\gamma$. As $p\to 1$, then $M_\gamma\to 1$; but very close to this limit,  $M_\gamma\approx 1$ is still a good approximation. For $p=1.01$, KS produces, $M_\gamma \approx 1.06$ (Fig. 7 and Table 1). This is an error of $6\%$ which is small considering that it is so close to the singular limit. An error of around $1\%$ away from $p=1$ is therefore reasonable. In the limit $p=3$ there is no detectible error in $M_\gamma$ from KS to three decimal places (Table 2).

KS is an established method used by many researchers in turbulent diffusion studies, as noted in the earlier references in this paper. However, some researchers, \cite{Thomson2005}, \cite{Nicolleau2011},  \cite{Eyink2013}, have expressed reservations regarding KS which they believe produces incorrect scalings for pair diffusion statistics because of the lack of true dynamical sweeping of the inertial scales eddies by the larger scales. These authors strongly support locality which they believe to be true in reality, and their inference that KS {\em mus}t be incorrect is a direct consequence of this thinking. As we are examining the hypothesis of locality itself here, such a presumption cannot be taken as a given.
Furthermore, \cite{Malik2015} has shown through an analysis of pairs of trajectories that the quantitative errors in KS due to the sweeping effect is in fact negligible provided that the simulations are carried out in a frame of reference moving with the large scale sweeping flow. The close agreement of KS with the geophysical data in Figure 1 obtained here provides further support for this conclusion.

\section{Discussion and conclusions}

Richardson pioneered the scientific discipline of turbulent particle pair diffusion, and assumed a locality scaling for the pair diffusivity due to the inclusion of one data-point from a non-turbulent context. Here, the reappraised 1926 data shows an unequivocal non-local scaling for the turbulent pair diffusivity, $K\sim \sigma_l^{1.564}$, Fig. 1. Consequently, the foundations of turbulent pair diffusion theory have been re-examined here in an effort to resolve one of the most important and enduring problems in turbulence. 

The main contribution of this investigation is to propose a new non-local theory of turbulent particle pair diffusion based upon the principle that the turbulent pair diffusion process in homogeneous turbulence  is governed by both local and non-local diffusional processes. The theory preditcs that in turbulence with generalized energy spectra, $E(k)\sim k^{-p}$, over extended inertial subranges, the pair diffusivity scales like, $K(p)\sim \sigma_l^{\gamma_p}$, with $(1+p)/2<\gamma_p\le 2$, in the range $1<p\le 3$, which is intermediate between the purely local and purely non-local scaling power laws. The reappraised 1926 geophysical data, Fig. 1, Table 1, provides strong support for the new theory.

Additional support comes from a Lagrangian simulation method, KS, whose results for turbulence with intermittency is in remarkably close agreement with the geophysical data in Figure 1, to within $1\%$ error in the scaling power for energy spectra in the accepted range of intermittency; for $E(k)\sim k^{-1.74}$, KS produces, $K_{\mu_I}\sim \sigma_l^{1.570}$, which is an error of just $0.4\%$ in the scaling power. All other predictions of the new theory have also been confirmed using KS.
Thus, an important corollary of the present work is that KS is a more accurate Lagrangian simulation method than previously thought. 

A detailed mathematical approach has been developed by expressing the pair diffusivity through a Fourier integral decomposition. {\em a priori} assumptions regarding locality was not made, and this has led to  an expression for $K$ as the sum of local and non-local contributions, equation (4.29). A feature of this approach is that it illustrates how various scalings and closure assumptions are inherent in developing any theory for pair diffusion. Such assumptions are always present even in apparently simpler scaling rules, but are often hidden and unstated.

This work also raises questions about previous works. Some DNS in particular have reported pair separation scaling that are close to, $\sim t^3$. How convincing is this in support of a locality hypothesis? 

In the first place, Richardson's 1926 data still remains the most comprehensive data on large geophysical scales till now, and it shows unequivocal non-local scaling. It is difficult to reconcile this with the apparent locality scaling reported in some studies, especially in view of the very short inertial subranges that they contain.

Furthermore, fully developed turbulence containing intermittency in a large inertial subrange should not in fact produce a R-O scaling of, $K\sim \sigma_l^{4/3}$, even under the locality hypothesis as we have seen in Section 5.4. The scaling in $\langle l^2\rangle $ in particular is very sensitive to the power in the energy spectrum through equation (41). With the intermittency observed in real turbulence and under the assumption of locality this would produce a scaling of, $K\sim \sigma_l^{1.35}$ to $\sim \sigma_l^{1.385}$, or equivalently $\langle l^2\rangle \sim t^{3.077}$ to $\sim t^{3.252}$, which should be detectable if present; but except for Richardson's revised 1926 dataset presented here, no study has reported a scaling greater than $\sim t^3$ strongly indicating that asymptotically large inertial subranges have not yet been achieved in experiments and DNS.

Under the current theory, we would expect even larger scaling powers, asymptoting approximately towards $K_{\mu_I}\sim \sigma_l^{1.57}$ and $\langle l^2\rangle_{\mu_I} \sim t^{4.65}$.

It is expected that attention will now turn towards the implications of this new theory for the general theory of turbulence, and for turbulence diffusion modeling strategies.\\

{\bf Acknowledgements} The author would like to thank SABIC for funding this work through project number SB101011, and the ITC Department at KFUPM for making available the High Performance Computing facility for this project. The author would also like to thank Mr. K. A. K. K. Daoud for producing the parallel version of the KS code. The author has benefitted from useful discussions on this topic with Professor A. Umran Dogan of King Fahd University of Petroleum \& Minerals and the University of Iowa.

\end{document}